\documentclass[twocolumn,showpacs,preprintnumbers,amsmath,amssymb,superscriptaddress]{revtex4-1}
 
\usepackage{graphicx}
\usepackage{dcolumn}
\usepackage{bm}
\usepackage{color}
\usepackage{todonotes}

\newcommand{\mups}{$~\mu$m$\cdot$s$^{-1}$}

\begin{document}

\title{Interaction of multiple particles with a solidification front : \\ from compacted particle layer to particle trapping
}

\author{Brice Saint-Michel}
\altaffiliation{Now at: ENS Lyon, Univ Claude Bernard, CNRS, Laboratoire de Physique, F-69342 Lyon, France}
\affiliation{Aix Marseille Univ, CNRS, Centrale Marseille, IRPHE, Marseille, France}

\author{Marc Georgelin}
\affiliation{Aix Marseille Univ, CNRS, Centrale Marseille, IRPHE, Marseille, France}

\author{Sylvain Deville}
\affiliation{Ceramic Synthesis and Functionalization Lab, UMR3080, CNRS/Saint-Gobain, Cavaillon, France}

\author{Alain Pocheau}
\affiliation{Aix Marseille Univ, CNRS, Centrale Marseille, IRPHE, Marseille, France}

\date{\today}

\begin{abstract}
The interaction of solidification fronts with objects such as particles, droplets, cells, or bubbles is a phenomenon with many natural and technological occurrences.
For an object facing the front, it may yield various fates, from trapping to rejection, with large implications regarding the solidification pattern. However, whereas most situations involve multiple particles interacting with each other and the front, attention has focused almost exclusively on the interaction of a single, isolated object with the front. Here we address experimentally the interaction of multiple particles with a solidification front by performing solidification experiments of a monodisperse particle suspension in a Hele-Shaw cell, with precise control of growth conditions and real-time visualization. We evidence the growth of a particle layer ahead of the front at a close-packing volume fraction and we document its steady state value at various solidification velocities. We then extend single particle models to the situation of multiple particles by taking into account the additional force induced on an entering particle by viscous friction in the compacted particle layer. By a force balance model, this provides an indirect measure of the repelling mean thermomolecular pressure over a particle entering the front. The presence of multiple particles is found to increase it following a reduction of the thickness of the thin liquid film that separates particles and front. We anticipate the findings reported here to provide a relevant basis to understand many complex solidification situations in geophysics, engineering, biology, or food engineering, where multiple objects interact with the front and control the resulting solidification patterns.
\end{abstract}

\maketitle

\section{Introduction}
The interaction of particles with a solidification front is a phenomenon encountered in numerous natural and technological situations, such as the evolution of frozen soils by frost heave or by ice lenses formation~ \cite{Corte1962,Zhu2000,Rempel2004,Dash2006,Peppin2013,Saruya2013,Anderson2014,Saruya2014}, cryo-biology~\cite{Bronstein1981,Korber1988}, food industry~ \cite{Velez-Ruiz2007} and materials science, from alloy casting in presence of particles~\cite{Stefanescu1988,Asthana1993,Liu2008} to the fabrication of bio-inspired composites~\cite{Deville2007a}. When a particle suspension freezes, a solidification front interacts with the dispersed particles by short range thermomolecular forces induced by Van der Waals like interactions \cite{Chernov1976,Wettlaufer2006}. Depending on the freezing conditions and characteristics of the suspension, this interaction may affect the particle distribution in the frozen phase in a number of ways and thus yield a large variety of segregation patterns, with important implications regarding the effective properties of the resulting solid~\cite{Zhang2005,Youssef2005,Deville2006c,Deville2009}. 

In most of the natural and technological occurrences of freezing suspensions, multiple particles interact with each other an with the front, sometimes forming ahead of the front a dense layer of particles referred to as a pile-up~\cite{Korber1985} or a build-up~\cite{Stefanescu1988} of particles. Whereas this layer may directly impact the trapping mechanism, little work has been done so far to understand the interaction of multiple particles with the front due to the inherent complexity of the problem. In particular, the trapping mechanism has long been studied by considering a single, isolated particle interacting with the freeze front, both experimentally~\cite{Cisse1971,Zubkho1973,Korber1985,Sen1997} and theoretically~\cite{Chernov1976,Chernov1977,Gilpin1980,Uhlmann1964,Hadji1999,Hadji2002,Azouni1998,Shangguan1992,Asthana1993,Catalina2000,Rempel1999,Rempel2001,Garvin2003,Wettlaufer2006,Park2006}. 

The objective of this study is to experimentally address the solidification regime of a particle suspension where multiple particles interact with the front, and use its characteristics to improve the modeling of multiple particles/front interaction.

For this, a unidirectional freezing experiment in a Hele-Shaw sample is performed using a monodisperse particle suspension. We focus on the steady state regime where a  
compacted particle layer of constant width $h$ is pushed at a constant velocity $V$ by the front. As the steady-state width $h$ of this layer is directly linked to the trapping mechanism, its evolution with the velocity $V$ provides a mean to document it and question its modeling in the presence of particles.

We thus build a mechanical model for trapping that takes into account the presence of multiple particles through the additional force exerted on an entering particle by the particle layer, following viscous  friction and pressure drop accross it. The particle layer width $h$ then provides
an indirect measure of the mean pressure $\bar{P_{\rm T}}$ exerted by thermomolecular forces on a particle entering the front. The elaboration of this model is facilitated by the absence of frozen fringes and thus by a homogeneous distribution of hydrodynamical resistance throughout the particle layer. Comparison of the resulting mean thermomolecular pressure to that predicted by existing models of single particle/front interaction questions both their relevance and their extension to multiple particles/front interaction.

Section \ref{Experiment} reports the experimental set-up and the main observations and measurements. The model that links the layer thickness $h$, the solidification velocity $V$ and the mean thermomolecular pressure on an entering particle $\bar{P_{\rm T}}$ is exposed in section \ref{Model}. The resulting indirect measure of this pressure is reported in section \ref{Results} together with a comparison to the value proposed by the single particle models. A discussion about quantitative deviations and the extension of single particle models to particle layers is drawn in Section \ref{Discussion}.

\section{Experiment}
\label{Experiment}

\subsection{Experimental set-up}

\begin{figure}[h]
\includegraphics[width=8.5cm]{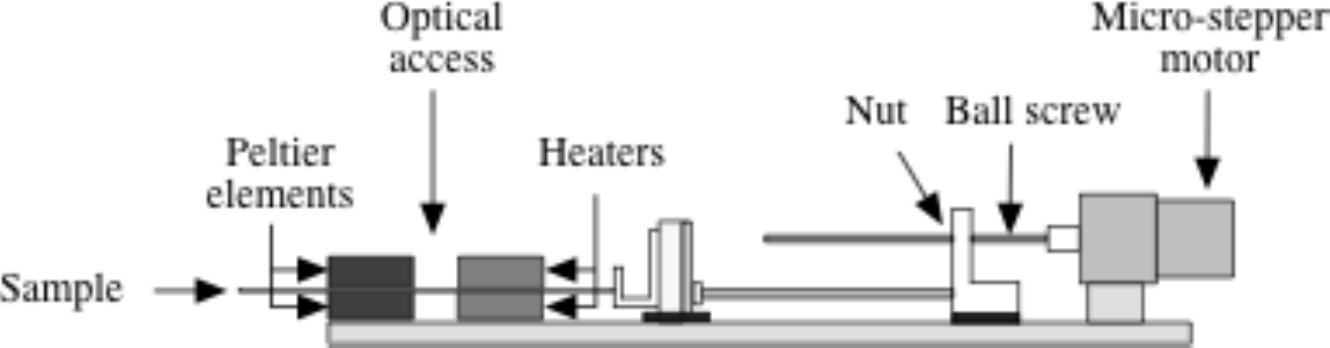}
\caption{Sketch of the experimental set-up. A thin sample sandwiched between heaters and coolers is pushed at a prescribed velocity by a micro-stepper motor coupled to a linear track. This allows solidification under controlled conditions together with a real-time non-invasive visualization.}
\label{Set-up}
\end{figure}

The experimental set-up has been originally designed to perform unidirectional solidification of binary solutions in thin samples~\cite{Georgelin1998,Pocheau1999}. In this setup, a sample is pushed at a controlled velocity $V$ in an imposed thermal gradient $G$, and video microscopy is performed (see Fig.~\ref{Set-up}) in the vicinity of the solidification interface to follow its evolution.

The thermal field is set by top and bottom heaters and coolers, electronically regulated to better than $0.1^{\circ}$C by resistive sheets and Peltier devices. Top heater and cooler are designed and fixed to provide altogether a planar surface on which the sample can slide, the thermal expansion of the top heater being anticipated. Bottom heater and cooler are then placed so as to sandwich samples, not by rigid fixation but by using springs, so as to reduce mechanical constraints and allow self-adjustment. An external circulation enables heat to be evacuated from the Peltier devices and also from the set-up sides to minimize thermal boundary perturbations.

The thermal gradient is set by both the gap between heaters and coolers and their temperature difference. Heaters and coolers temperatures are chosen so as to place the melting isotherm in the center of the gap, both to achieve an easier visualization of the solidification interface and to minimize of the thermal gradient dependence on the sample velocity $V$~\cite{Georgelin1998,Pocheau2009}.
The gap was $10$ mm and the imposed temperatures $20^{\circ}$C and $-20^{\circ}$C for heaters and coolers and $-10^{\circ}$C for the circulating external flow. A closed dry atmosphere helped avoid condensation and ice formation.

Sample displacement is provided by a pushing device held on a linear track (see Fig.~\ref{Set-up}). Regular translation of the pusher is produced by a recirculating ball screw mounted on a screw rotated at a controlled rate by a microstepper motor. 
Vibration sources from the motor rotation are minimized by the large number of steps (200 per turn) and of microsteps (32 per step), together with the generation of Foucault currents to slow rotation at the end of microsteps. This provides 6400 microsteps per turn which, with a $5$~mm screw pitch, yields an elementary displacement of $0.8~\mu$m. Modulations of sample velocities on a screw turn reduced to a sinusoidal part with a relative magnitude $\delta V/V$ less than $3 \%$ at most. Regular solidification velocities could thus be applied up to $50$\mups{}.

Visualization of a part of the solidification front is achieved by a CCD camera with time-lapse recording. To minimize perturbations in the vicinity of the sample, large frontal distances have been preferred. For this an exploded optical chain with a $50$~mm focal lens has been used instead of a microscope. Magnification could allow for the visualization of fields extended from $200~\mu$m to $2$~mm with a resolution limited by diffraction to about $0.2~\mu$m. As the camera involved $1024 \times 768$ pixels, the numerical resolution was thus in practice above the optical resolution.

Two kinds of observations could be performed: in transmission or in reflection. In transmission, the suspension and the frozen phase appear grey
but the particle layer is opaque, which forbids the observation of its constituents. In reflection, the suspension and the frozen phase still appear grey
but the particle layer is bright, which allows particles close to the sample surface to be individually followed in real time. As in both cases, the particle layer width was the same, we preferred the observation by reflection to provide access to some particle dynamics. In addition, some post-mortem complementary observations were made by confocal microscopy (Leica SP8, combined to a Leica DM6000 optical microscope) to better visualize particle ordering in the vicinity of a plate. 
 
Samples were made of glass plates sandwiching the suspension. Their dimensions (top glass $100\times 45 \times 0.7$~mm$^3$ ; bottom glass $150\times 50 \times 0.8$~mm$^3$) have been taken long and large enough to allow a large central zone ($>35$~mm) free of boundary effects. They were separated by calibrated spacers made of polypropylene sheets. After having filled the samples with the suspension by capillarity, they were sealed with cyanoacrylate and epoxy glue.

For binary solutions, the sample depth was usually chosen close to the natural width of microstructures, about $100 \mu$m~\cite{Georgelin1998,Pocheau1999,Deschamps2008}. This enabled a single layer of microstructures to settle in the sample, thereby facilitating visualization. Implications of confinement could be studied on dendritic sidebranching by experiment~\cite{Georgelin1998} or on cell forms by 3d phase field simulations~\cite{Gurevich2010,Ghmadh2014}.
Both confirmed that microstructure layers grown in a thin, but thick enough, sample were physically equivalent to those growing in an unconfined environment. These findings led us to choose the sample depth large enough compared to both the typical sizes of ice microstructures and the particle diameter. In addition, as multi-layers of microstructures were allowed here, a sample depth of $125~\mu$m has been adopted.

The suspensions, manufactured by Magsphere Inc., contained monodisperse polystyrene spheres of $3~\mu$m diameter and density $1.03$,  at volume fraction $\phi_0=10\%$ or $20\%$. Sphericity and monodispersity were confirmed by confocal microscopy. The suspensions were stable over months.
Although they might sediment in the samples, the particles stayed disperse enough to be able to redistribute in the sample depth when encountering the solidification front and its adjacent particle layer. Solutal effects were investigated by filtering out the particles using chromatography micro-filters. They led here to a critical velocity of about  $10$\mups{} for the morphological instability of planar fronts.

The physical parameters of the experiment are given in Tab.~\ref{tab:parameters}.

\begin{table}
	\begin{tabular}{l c r l}
    \hline
    Name\rule{0pt}{13pt} 					& Symbol		& Magnitude				& Unit	\\
    \hline
    Dynamical viscosity\rule{0pt}{13pt} 	& $\mu$			& $1.8\times 10^{-3}$	&	Pa$\cdot$s	\\
  Volumic latent heat					& $L_v$			& $3.3\times 10^{8}$	& 	J$\cdot$m$^{-3}$ \\
    Melting point							& $T_{\rm M}$	& $273.15$				&	K	\\
    Surface tension (with ice)				& $\sigma$		& $2.9\times 10^{-2}$	& 	J$\cdot$m$^{-2}$ \\
    Hamaker constant (water/PS)				& $\mathcal{A}$	& $1.3\times 10^{-20}$	&   J \\
    Thermal conductivity (water)			& $k_{\rm m}$	& $0.561$				&	W$\cdot$K$\cdot$m$^{-1}$ \\
    Thermal conductivity (PS)				& $k_{\rm p}$	& $0.115$				&	W$\cdot$K$\cdot$m$^{-1}$ \\
    Particle radius							& $R$			& $1.5 \times 10^{-6}$	&	m	\\
    Thermal gradient						& $G$			& $4\times 10^3$		& 	K$\cdot$m$^{-1}$ \\
    \hline
    \end{tabular}
    \caption{The physical quantities used throughout the manuscript.}
    \label{tab:parameters}
\end{table}

\vskip\baselineskip	
\subsection{Observations and measurements}
\label{ObservationsMeasurements}

This study focuses on the velocity range $0.5 \leq V \leq 20$\mups with a special attention to velocities below the critical velocity of a planar front (around $10$\mups). Therefore, most data will refer to a planar solidification interface. Other velocities and/or particle diameters yield different solidification patterns that will be reported in a forthcoming paper.

\begin{figure}[h!]
\includegraphics[width=8.6cm]{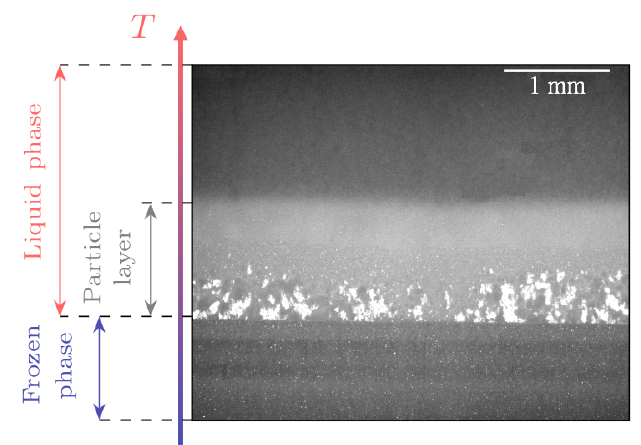}
\caption{Steady state at $\phi_0 = 20\%$ and $V=3$\mups{} showing a dark frozen phase (bottom) followed by a grey particle layer (middle) and the liquid suspension (top). The frozen phase shows striations reminiscent of ice lenses. The particle layer displays bright patches corresponding to an ordering of the particles in the vicinity of the plates.
}
\label{SteadyStateV=3}
\end{figure}

A typical steady-state, obtained at $V=3$\mups, is displayed in Fig.~\ref{SteadyStateV=3}. We observe a grey and bright layer surrounded by the frozen phase (bottom) and the liquid phase (top). The layers are separated by sharp boundaries.
The frozen phase shows repetitive black striations which refer to a phenomenon that is both time-dependent and uniform along the front. This is reminiscent of the segregation provided by the formation of ice lenses~\cite{Anderson2012}, but a more detailed analysis will be required to state this firmly. 
However, no implication of these striations could be noticed on the rest of the system, especially on the front shape or the width and the organization of the adjacent bright layer. We shall therefore ignore them here.

\begin{figure}[h!]
\includegraphics[width=8.6cm]{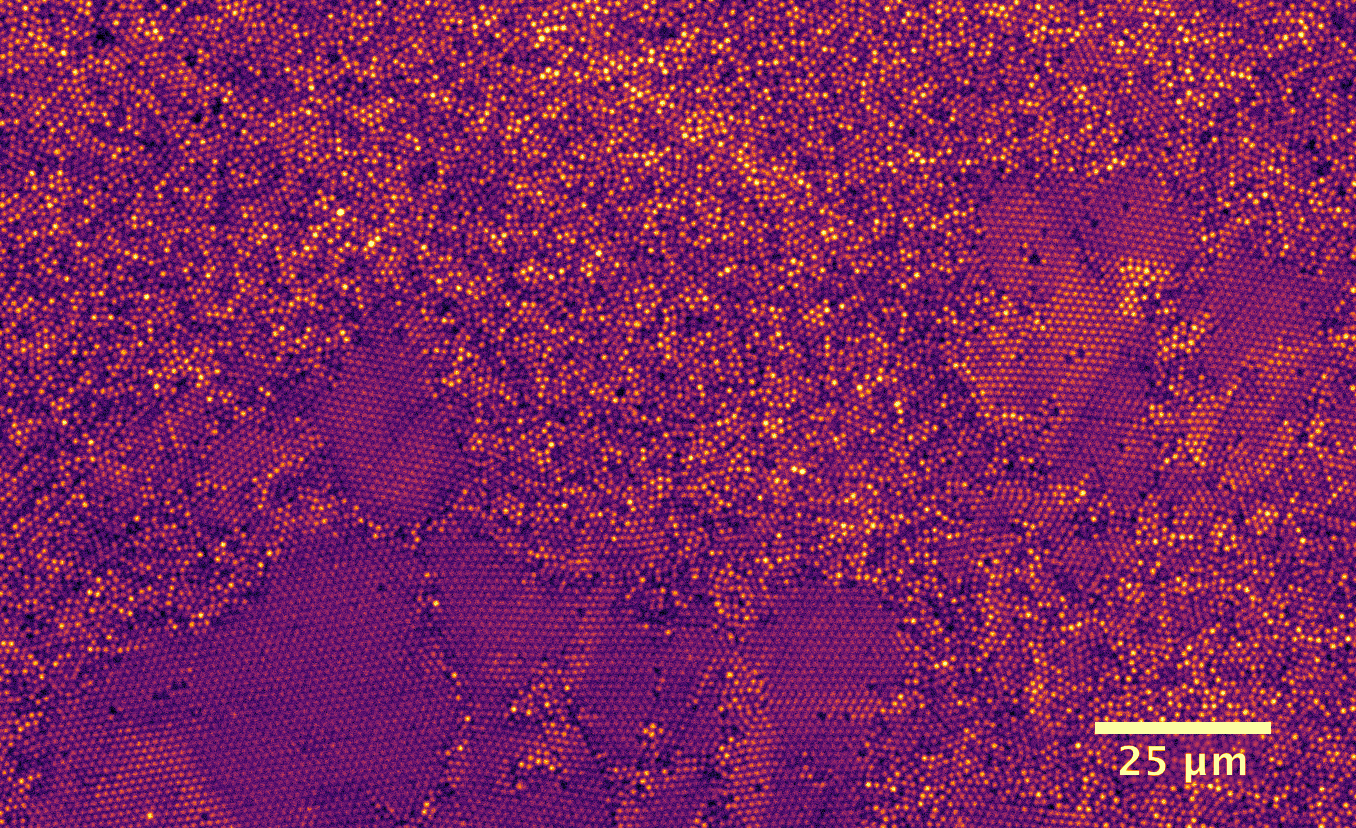}
\caption{Confocal microscopy image of the compacted layer of particles close to a sample plate. The particle diameter is $1~\mu$m. The particle layer shows close-packed arrangements of particles into both disordered regions or colloidal crystals of relatively large size. This noticeable geometrical order is presumably imposed by the planar glass surface.}
\label{Confocal}
\end{figure}

The layer between the frozen phase and the liquid phase corresponds to the compacted layer of particles
studied by Anderson and Worster~\cite{Anderson2012,Anderson2014}. Its existence indicates that the velocity at which particles enter the frozen phase is smaller than the growth velocity $V$, thereby requiring an increased particle concentration $\phi$ in steady-states to satisfy the particle flux balance in the whole system. This agrees with direct observation from confocal microscopy (Fig.~\ref{Confocal}) which reveals a close-packing of particles.

In the vicinity of the front, the particle layer displays bright patches with a broad size distribution. As revealed by confocal microscopy, they correspond to ordered clusters of particles near the plates (see Fig.~\ref{Confocal}) which likely result from the long-range ordering induced by the planar glass surface. The observations reveal that they evolve with time in both shape and size with possible internal reorganization responsible for cluster blinking (see Movie S1, Supporting Information). 
These clusters, reminiscent of particle crystallization in suspensions~\cite{Pieranski1983}, are thus likely limited to the vicinity of plates. They therefore correspond to a visual artifact induced by the plate with no prejudice on the remaining particle layer, both in width and depth.

The cluster brightness could be related to particles. It thus makes a definite difference with the frozen fringes observed by Anderson and Worster on smaller particles~\cite{Anderson2012}. In particular, their non-uniformity, their time-dependence and their disconnection with the front refer to features that do not fit with the current description of the frozen fringes. In addition, no steady bright zones similar to frozen fringes could be observed close to the front. This supports the absence of frozen fringes in this experiment.

\begin{figure*}[ht!]
\centering
\includegraphics[width=18cm]{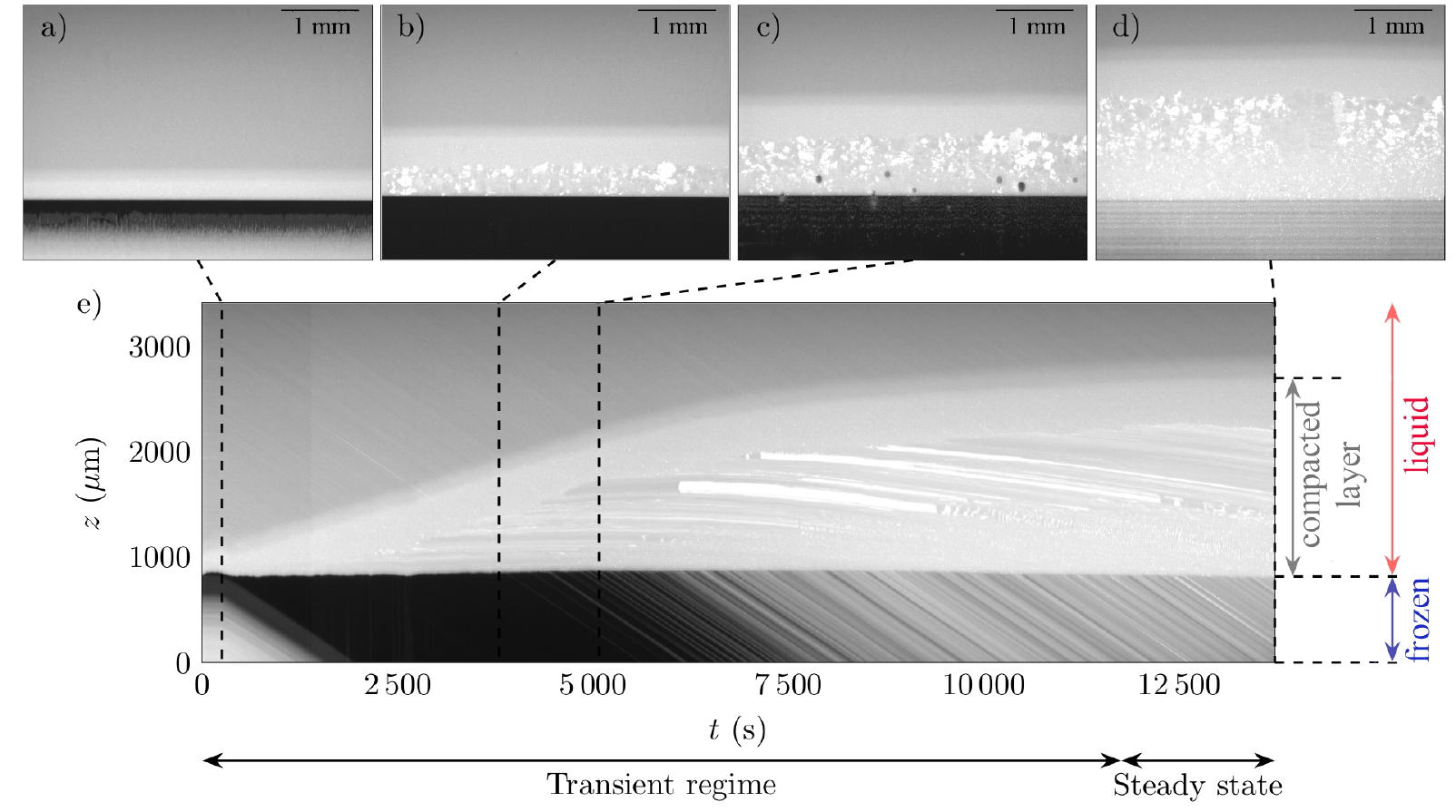}
\caption{Initial stage of solidification : $\phi_0 = 20\%$, $V=0.5$\mups{}. The corresponding movie is displayed in Supporting Information, Movie S1. (a) - (d) Snapshots taken respectively $220$, $3789$, $5062$ and $13720$ s after beginning of solidification. They show, from top to bottom, the suspension (grey zone), the particle layer (bright zone) and the solid phase (dark when no particles are trapped (a to c), grey when some are (d)).
(e) Spatio-temporal diagram provided by the juxtaposition at increasing times of a cut of images taken at a fixed position. Space is reported on ordinates over 768 pixels corresponding to $3428~\mu$m. Time is reported on abscissa over $1372$ pixels corresponding to $13720$~s. The times corresponding to images (a) to (d) are indicated on the top of the diagram. The common slope of the bright lines displayed in the solid phase refers to the advection of embedded particles at sample velocity $V$. The diagram shows that particles are first repelled, as revealed by the dark zone behind the front. Meanwhile, the particle layer thickness increases until the particle capture by the solidification front sets in (bright lines in the solid phase). The particle layer then reaches a steady state and a constant thickness.
}
\label{DST}
\end{figure*}

Starting from the beginning of solidification, the particle layer grows linearly as displayed in fig.~\ref{DST} in a series of snapshots and in a spatio-temporal diagram (see Movie S1, Supporting Information). Meanwhile, the frozen phase behind the front appears dark, meaning that no particle enters it, at least close to the sample plates. This is confirmed by transmission observations, where the brightness of the frozen phase proves that it contains no particle in the whole sample depth. 
As the incoming particle flux in the particle layer is $V \phi_0$, particle conservation yields $V \phi_0 = (\phi-\phi_0) {\rm d}h/{\rm d}t$ where $h$ denotes the layer thickness, $\phi_0$ the particle concentration in the suspension and $\phi$ its mean value in the particle layer. The linear increase of $h$ with $t$ thus enables an indirect measure of the particle concentration $\phi$ in the layer. The fit of its growth rate yields $\phi=0.634  \pm 0.007$, close to the random close-packing value $\phi_{\rm p}=0.637$ and in agreement with observation of clusters of particles in contact with each other (Fig.~\ref{Confocal}). 
This determination supports previous assumptions~\cite{Anderson2012,Anderson2014} and predictions~\cite{Peppin2006} that a layer of particles under such conditions is at the close-packing concentration $\phi_{\rm p}$. 
It thus confirms that the denomination "compacted layer of particles" is relevant.
The growth of the layer ends at a definite layer thickness which is thus representative of the suspension and solidification conditions.
As only steady states will be addressed from now on, we shall then still denote their permanent layer thickness $h$ in the remaining. This stop of growth means that the entrance flux of particles in the frozen phase now equals their incoming flux $V \phi_0$ in the particle layer. It thus indicates that the steady state layer width $h$ is intimately linked to the repelling/trapping mechanism of particles by the front. In addition, as it is simple and accurate to measure, it thus stands as an ideal quantity to indirectly probe the particle/front interaction mechanism in this experiment.

\begin{figure*}[ht!!]
\includegraphics[width=16cm]{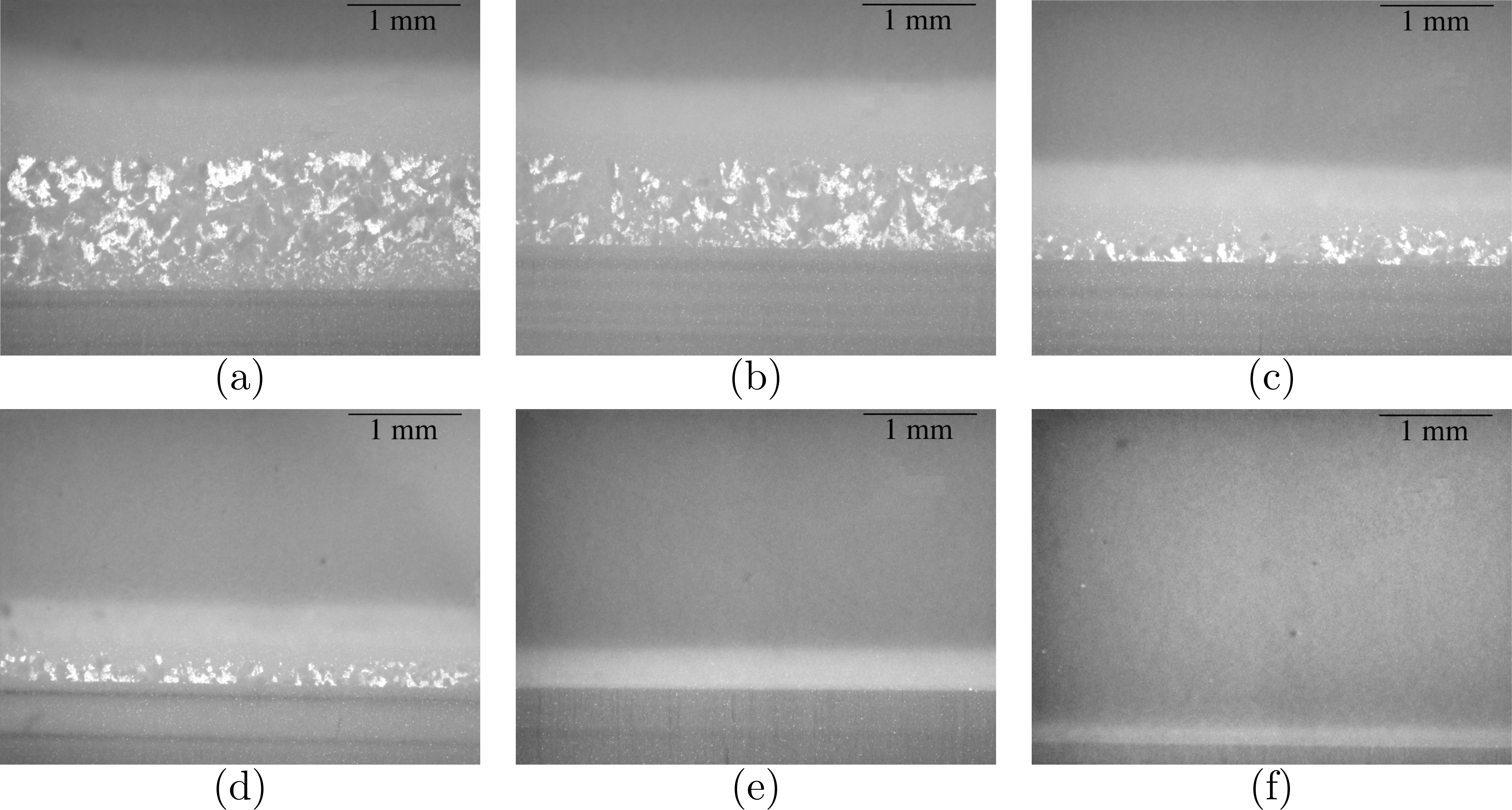}
\caption{Particle layer (grey zone) in steady states at $\phi_0 = 20\%$ and various front velocities $V$ : $1$, $2$, $4$, $5$, $10$ and $20$\mups{}. The layer thickness decreases with the front velocity. Below $V=10$\mups, it shows white patches which correspond to clusters of particles on the sample plates (a, b, c, d). 
As the location of the images is arbitrary, the variation of the front position within the observation frame is meaningless.
}
\label{SteadyStates}
 \end{figure*}

Fig.~\ref{SteadyStates} displays steady states obtained at various solidification velocities $V$. They all show a similar kind of patterns with a particle layer and a frozen phase possibly displaying, at low velocities, bright patches and striations respectively. The layer thickness $h$ shows a continuous decrease with the velocity $V$. This is documented for two suspension concentrations $\phi_0=0.1$ and $0.2$ in fig.~\ref{G,h,V}, where no dramatic change in the evolution of $h$ can be noticed for the two data points close or above the critical velocity of destabilization of planar interface, about $10$\mups here. 

The following sections of this article are devoted to understand both this trend and its order of magnitude so as to gain relevant insights on the underlying repelling/trapping mechanism of particles by the solidification front.

\begin{figure}[ht!]
\includegraphics[width=8cm]{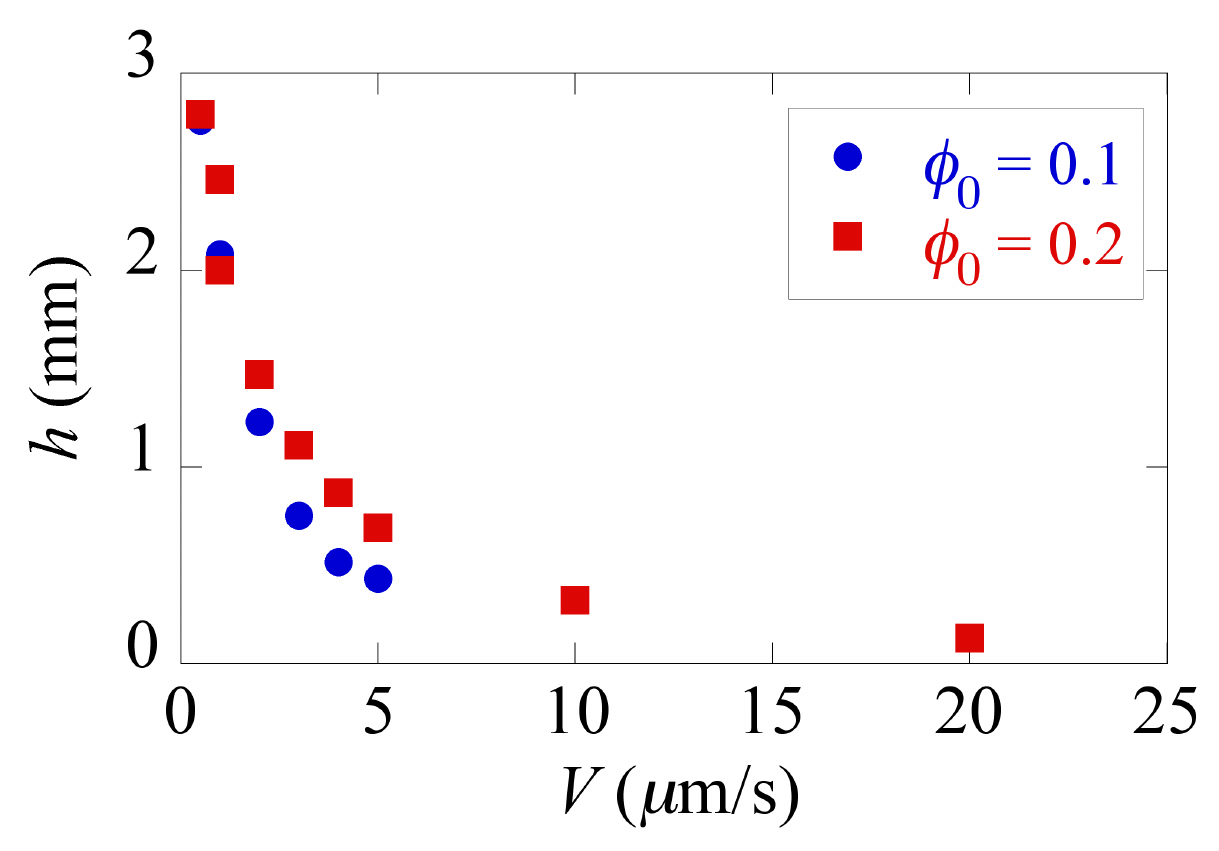}
\caption{Evolution of the steady state thickness $h$ of the particle layer with the front velocity $V$ at particle concentrations of the suspension $\phi_0=0.1$ and $0.2$.
}
\label{G,h,V}
\end{figure}

\section{Mechanical modeling}
\label{Model}

The particle layer refers to a self-organized steady state in which the repelling/trapping mechanism plays a key role. To access relevant information on it, we model the mechanics of the particle layer, especially viewed from the particles that enter the front. As in steady state the forces applied on the particle equilibrate, the magnitude of the short-range thermomolecular force could then be deduced from the remaining forces.

Three forces apply on a particle close to the solidification front (see Fig.~\ref{ForcesParticuleEntrante}) :

i) The thermomolecular force $\mathbf{F}_{\rm T}$ exerted by the solidification front on an entering particle. It results from Van der Waals interactions between the particle and the front, through the fluid.

ii) The lubrication force $\mathbf{F}_{\rm l}$ on an entering particle. It results from viscous effects in the thin film separating the particle from the front.

iii) The force $\mathbf{F}_\mu$ exerted by the particle layer on an entering particle. It results from pressure drop and viscous stress induced by the fluid flowing across the particle matrix. It is transmitted to an entering particle by contacts along the particle matrix.

Whereas the two former forces push the particle toward the front, the latter repels it. Their resultant will be labeled $\mathbf{F}$. These forces are addressed in more detail below before we can deduce the thermomolecular force from mechanical equilibrium.

\subsection{Mechanical balance}
\label{MechanicalBalance}

\begin{figure}[h!]
\includegraphics[width=6cm]{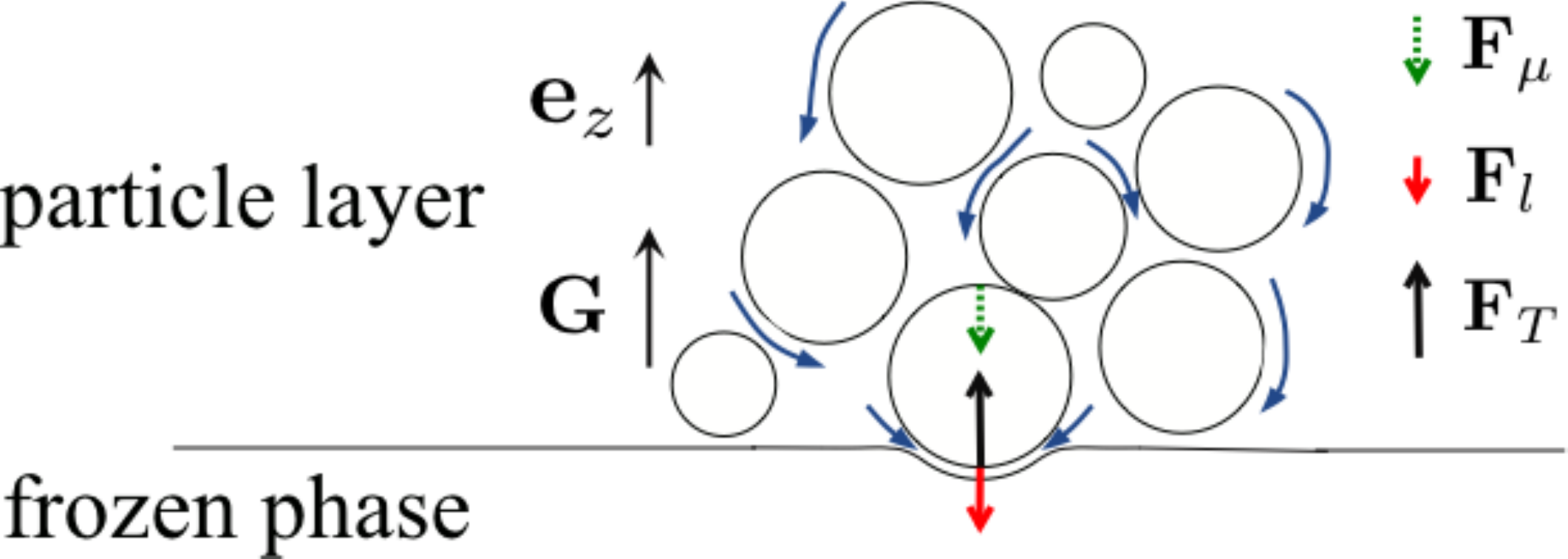}
\caption{Sketch of the forces acting on a particle entering the frozen phase: a thermomolecular repelling force $\mathbf{F}_T$ exerted by the solidification front (black arrow); a lubrication force $\mathbf{F}_l$ induced by the liquid flow in the thin film separating the particle and the front (red arrow); a force $\mathbf{F}_\mu$ exerted by the particle layer on the particle, as a result of the viscous friction induced by a liquid flow through the particle matrix (green dashed arrow). Blue curved arrows symbolize the liquid flow; $\mathbf{G}$ recalls the thermal gradient and $\mathbf{e}_z$ the direction of $z$-axis. As particles are randomly distributed, their section by a plane displays different radii and few contact points here, although they actually share the same radius and are in contact with each other.
}
\label{ForcesParticuleEntrante}
\end{figure}

For convenience, we shall consider the average over a particle section $\pi R^2$ of the $z$-components of these three forces and their resultant on the thermal gradient direction $\mathbf{e}_z=\mathbf{G}/G$.
The corresponding pressures will be referred to as $\bar{P_\mu}$, $\bar{P_{\rm l}}$, $\bar{P_{\rm T}}$ and $\bar{P}$ respectively.
They thus satisfy the relationship $\bar{P} = \bar{P_\mu}+ \bar{P_{\rm l}} +\bar{P_{\rm T}}$, where $\bar{P_\mu}<0$, $\bar{P_{\rm l}}<0$ and $\bar{P_{\rm T}}>0$ involve the following expressions :
\begin{enumerate}

\item Viscous force ${\bf F_{\mu}}$, $\bar{P_\mu}$

In steady states, mass conservation of fluid and of particles yields the following mean velocities for the fluid, $v_{\rm f}$, and the particles, $v_{\rm p}$, in the front frame: $v_{\rm f} = -  V (1-\phi_0)/(1-\phi)$, $v_{\rm p} = -  V \phi_0/\phi$  (Fig.~\ref{Suspension}). As they differ, viscous stresses are generated on the particle matrix. Regarding the fluid, they result in a pressure drop $\Delta p$ across the particle layer, referred to as cryosuction.

From a mechanical point of view, this pressure drop expresses the rate of momentum flux on the direction $\mathbf{e}_z$ that is lost by the fluid and gained by the particle matrix, by viscous stresses on the particle surfaces. As the particle layer is close-packed,
 this momentum is entirely transmitted to the entering particles and results in a mean pressure $\bar{P_\mu}$ on each of them. However, as particles occupy only a part $\phi$ of a suspension section, the mean pressure on an entering particle is larger than the pressure drop and reads $\bar{P_\mu} = \Delta p / \phi$. In particular, its excess with respect to the pressure drop refers to the viscous stresses that are also undergone by the particle layer, in addition to the viscous drop itself.

Regarding hydrodynamics, the compacted layer of particles behaves like a porous medium.
In addition, following the absence of frozen fringes (Sect.~\ref{ObservationsMeasurements}), it may be considered as homogeneous. Accordingly, its internal viscous dissipation may be modeled by Darcy's equation, $\mu U = - k {\rm d}P/{\rm d}z$, where $\mu$ denotes the liquid viscosity, $k$ the medium permeability and $U$ 
the flux of liquid volume in the $z$-direction, viewed in the frame of the particle matrix. Taking into account the mean velocity difference $v_{\rm f} - v_{\rm p}$ between the liquid phase and the particles and the part $(1-\phi)$ of a suspension section occupied by the liquid, one gets $U = (1-\phi) (v_{\rm f} - v_{\rm p})$ and $U = - V (\phi-\phi_0)/\phi$.

Following the indirect measure reported in Sec.~\ref{ObservationsMeasurements}, the particle concentration $\phi$ in the compacted layer of particles
will be taken as that of random close packing : $\phi_{\rm p}=0.64$. As the suspension concentration $\phi_0$ is smaller than $\phi_{\rm p}$, one notices that $U$ is negative, which means that liquid flows from the suspension towards the front. On the other hand, as the suspension is monodisperse and random, its permeability may be modeled by the Kozeny-Carman relation: $k(\phi,R)= (1-\phi)^3/\phi^2 \, R^2/45$. This yields a pressure drop across the particle layer $\Delta p = \mu U h / k(\phi,R)$ and finally a mean extra pressure on the entering particles $\bar{P_\mu} =  - \mu |U| h / [ \phi k(\phi,R) ]$.

\begin{figure}[h!]
\includegraphics[width=6cm]{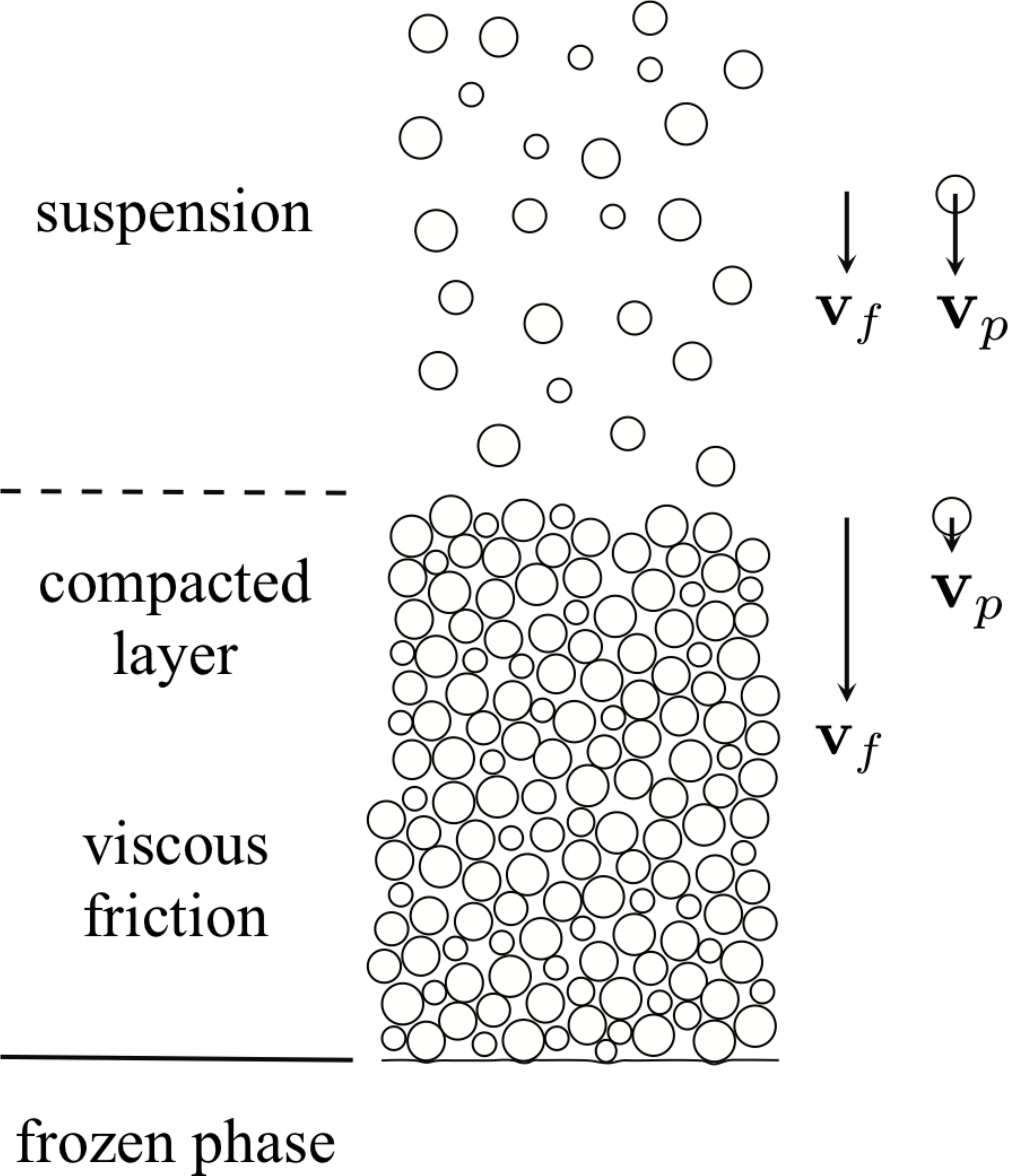}
\caption{Sketch of the system that shows the suspension, the compacted layer of particles and the frozen phase. The mean velocities of the fluid and of the particles with respect to the solidification front are denoted $\mathbf{v}_{\rm f}$ and $\mathbf{v}_{\rm p}$ respectively. In the suspension, they are both equal to the opposite of the solidification velocity. However, in the compacted layer of particles, they differ following the rise of particle volume fraction, thus resulting in viscous friction. As particles are randomly distributed, their section by a plane displays different radii although they 
actually share the same radius.
}
\label{Suspension}
\end{figure}

\item Lubrication force ${\bf F_{\rm l}}$, $\bar{P_{\rm l}}$

When a particle is displaced with respect to the suspension, some fluid has to fill the freed volume. For particles adjacent to the front, such a flow occurs in the film that separates them from the front (Fig.~\ref{ForcesParticuleEntrante}). 
Its thinness then induces, by viscous friction, a noticeable pressure drop at the base of particles which results in an attractive force between them and the front. Referred to the cross-section of particles, it corresponds to a mean pressure $\bar{P_{\rm l}}$ on them.

Computation of $\bar{P_{\rm l}}$ may be performed in the framework of Stokes flow and of the lubrication approximation. The former yields this mean pressure to be proportional to the velocity $V_p$ of particles with respect to the suspension : $\bar{P_{\rm l}}= - f V_p$. The latter approximation yields the prefactor $f$ to be determined with respect to the geometry of the film, i.e. the front shape and the position of the particle with respect to it.

In the following and in most models, attention is focused on steady states of particles adjacent to the front for which the relative velocity $V_p$ with respect to the suspension is constant. However, we notice that the value of this velocity differs depending on whether the entering particle is isolated and repelled by the front or immersed in a steady particle layer and thus eventually trapped by the front. The former case is usually considered by models and yields $V_p=V$. The latter case is the one which applies here. It corresponds to $V_p=v_p + V$ or, equivalently $V_p = - U = |U|$. 

\item Thermomolecular force ${\bf F}_{\rm T}$, $\bar{P_{\rm T}}$

The thermomolecular force results from Van der Waals like interactions between the front and a particle. It is normal to the particle surface and inversely proportional to the cube of the distance $d$ between a surface point and the front. It thus gives rise to a thermomolecular pressure $P_{\rm T}=\mathcal{A}/6\pi \, d^{-3}$ where the Hamaker constant $\mathcal{A}$ depends on materials and solute.

Given the front shape and the position of the particle with respect to the front, the thermomolecular pressure may be integrated along the particle surface to give the mean thermomolecular pressure $\bar{P_{\rm T}}$.

 \item Thermodynamical equilibrium
  
To compute both the lubrication force and the thermomolecular force, the front shape must be determined for a given position of the particle with respect to it. It follows the Gibbs-Thomson relation:
\begin{equation}
\label{GT}
L_v \; t = P_{\rm T} + \sigma \kappa
\end{equation}
where the pressure melting term induced by the difference of density between water and ice has been neglected. Here $P_{\rm T}$, $\sigma$, $\kappa$ and $t= (T_{\rm M}- T_{\rm I})/ T_{\rm M}$ denote the thermomolecular pressure, the surface tension and curvature of the interface, and its relative undercooling with $(T_{\rm M},P_{\rm M})$ the temperature and pressure of a reference state and $T_{\rm I}$ the solid/liquid interface temperature.

In a thermal gradient where $T_{\rm I}\equiv G z_{\rm I}$, this provides a differential equation for the location $z_{\rm I}$ of the interface whose integration yields  the front geometry, given the position  of the particle with respect to the melting isotherm. Usually, numerical integration is required to determine the front shape and thus the channel geometry, especially when capillary effects are handled \cite{Rempel1999,Park2006}. 

\end{enumerate}

\subsection{Mechanical equilibrium}
\label{MechanicalEquilibrium}
  
Mechanical equilibrium reads: $\bar{P_\mu}+ \bar{P_{\rm l}} +\bar{P_{\rm T}} = 0$. It applies to particles involving a constant velocity $V_p$, either because they are repelled by the front or because they belong to a steady particle layer displaying continuous trapping. The present experiment refers to the latter case for which $V_p=|U|$ and thus $\bar{P_{\rm l}} = - f |U|$. Introducing $U_c= \bar{P_{\rm T}} / f$, the critical velocity for a single particle (i.e. for $h=0$), the mechanical equilibrium of an entering particle then reads:
\begin{equation}
\label{h,1/U}
h  \, \frac{ \mu}{\phi k} = \bar{P_{\rm T}} \, (\frac{1}{|U|} - \frac{1}{U_c})
\end{equation}
where we recall that the permeability $k$ depends on $\phi$ and $R$. This allows the determination of both the mean thermomolecular pressure $\bar{P_{\rm T}}$ at the trapping transition and the critical velocity $U_c$ from the data set of steady-state thicknesses $h$ at various velocities $U$.

In relation (\ref{h,1/U}), we notice that the left hand side depends on the suspension properties $\phi$ and $R$ whereas the right hand side seems at first glance to be independent of them. This difference traces back to physics of the repelling/trapping mechanism : part of it (the viscous pressure $\bar{P_\mu}$) refers to the suspension whereas the remaining part (lubrication and thermomolecular forces $\bar{P_{\rm l}}$ and $\bar{P_{\rm T}} $) refers to the interaction of an entering particle with the front. Thus, whereas the former deals with the whole particle layer, the latter corresponds to short range forces exerted in the very vicinity of an entering particle and formally decoupled from the rest of the suspension. Following this picture, the repelling/trapping mechanism is expected to remain the same, irrespective of the presence of a particle layer, so that $\bar{P_{\rm T}} $ and $U_c$ should be viewed as independent of $h$.
However, we shall show later on, in section \ref{ParticleLayerEffect}, that this corresponds to a first order approximation with respect to the particle layer thickness beyond which there exists an indirect, hidden, link between the trapping mechanism and the suspension conditions.

\section{Results and mean thermomolecular pressure}
\label{Results}

We first report the evolution of the width $h$ of the particle layer with respect to the relative velocity $U$. This will provide an indirect measure of the mean thermomolecular pressure $\bar{P_{\rm T}}$. To interpret its magnitude, we then revisit the Rempel-Worster model~\cite{Rempel2001} that refers to the interaction and the capture of a single particle by the front.

\subsection{Particle layer thickness and mean thermomolecular pressure}
\label{results,PT}
Figure \ref{G,h,1-U} reports the particle layer widths $h$ with respect to the inverse relative velocity  $1/|U|$.
It shows an affine trend beyond $U\approx 1$\mups{} followed at lower velocities by a departure towards a saturation. As the latter regime is confined to low velocities, it is much less representative of the suspension solidification than the former regime which englobes all velocities beyond.
We shall therefore skip it in the following and postpone its analysis to a forthcoming paper.

Above $U\approx 1$\mups, the affine evolution (\ref{h,1/U}) followed by data supports the mechanical modeling involving viscous forces.
In addition, the use of the relative velocity $|U|$ here instead of $V$ as in figure (\ref{G,h,V}) makes the  the two data sets $\phi_0 = 0.1$ and 0.2 collapse on the same master curve. In particular, whereas their fitted slopes agree to within $5 \%$ here, they would differ by about $30 \%$ if $V$ was used instead.
As $U$ corresponds to the relative velocity of the fluid with respect to particles, this supports the hydrodynamic origin of the forces which counterbalance the thermomolecular force.
 
The slope and the intersection with the $x$-axis displayed in Figure~\ref{G,h,1-U} respectively provide estimates of the mean thermomolecular pressure $\bar{P_{\rm T}}$ and of the critical velocity $U_c$.
However whereas the former determination is accurate, the latter is largely sensitive to the distribution and the uncertainty of data so that its determination can only be quite approximate. Data yield $\bar{P_{\rm T}} = 1000 \pm 63$~Pa and an extremely large and imprecise critical velocity $U_c$ which must thus be overlooked.

The value found for $\bar{P_{\rm T}}$, a kPa,stands two orders of magnitude below the thermomolecular pressure $P_{\rm T0}$ at the base of the particle evaluated for a usual expectation of a nanometer thick film: $P_{\rm T0}=\mathcal{A}/6\pi d_0^{-3} = 690$~kPa. To clarify this large difference, we turn attention towards the modeling of the repelling/trapping mechanism proposed by Rempel and Worster~\cite{Rempel2001} and question its prediction regarding the mean thermomolecular pressure $\bar{P_{\rm T}}$ on an entering particle.

\begin{figure}[h]
\includegraphics[width=8cm]{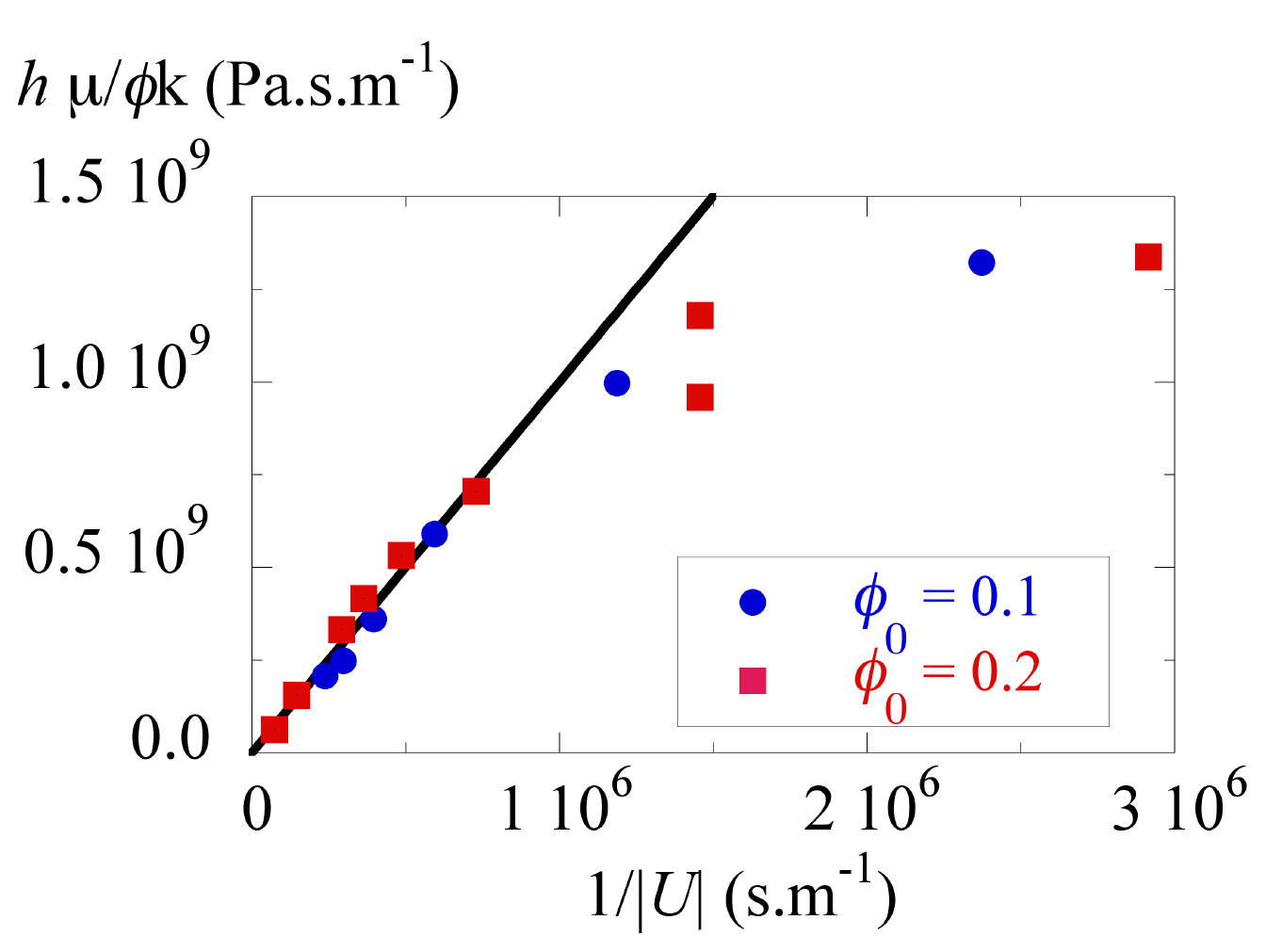}
\caption{Evolution of the particle layer thickness $h$ with $|U|$, the relative velocity of the fluid with respect to particles. The axes refer to the variables $\mu h/\phi k$ and $1/|U|$ to express relation (\ref{h,1/U}) by an affine relation with slope $\bar{P_{\rm T}}$.
}
\label{G,h,1-U}
\end{figure}

\subsection{Single particle modeling and thermomolecular constraint at critical state}
\label{SingleParticleCriticalState}

\begin{figure}[h!]
\includegraphics[width=6cm]{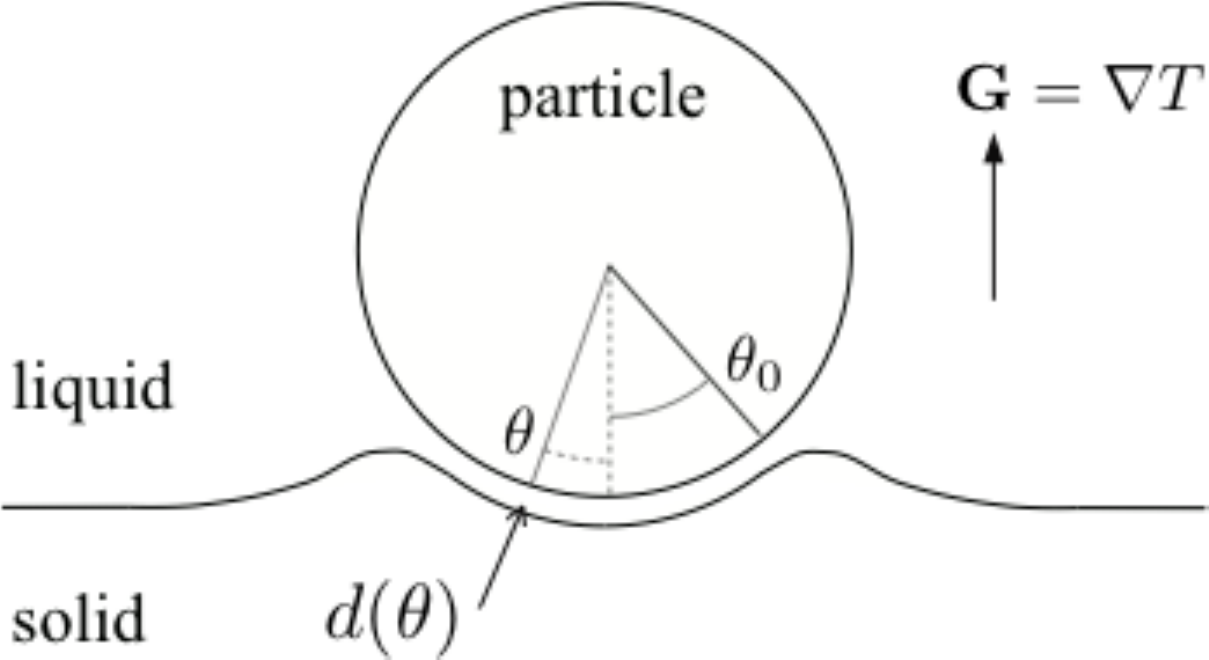}
\caption{Schematic diagram of a particle adjacent to the interface in the constant film thickness assumption where $d(\theta)\approx d_0$ for $\theta \leq \theta_0$. }
\label{SketchParticle}
\end{figure}

The most recent modelings of the interaction between a single spherical particle and the solidification front~\cite{Rempel1999,Rempel2001,Park2006} provide the following integral expressions of the $z$-component of the forces $F_{\rm l}$ and $F_{\rm T}$ on a particle :
\begin{align}
\label{Fl}
F_l & =  2\pi R^2  \, 6\mu R^2 V_p  \int_0^{\frac{\pi}{2}} \sin(\theta) \cos(\theta) \left [ \int_{\frac{\pi}{2}}^\theta \frac{\sin(\phi)} {d^3(\phi)} {\rm d}\phi \right ] {\rm d}\theta \\
\label{FT}
F_T & =  2\pi R^2 \, \frac{\mathcal{A}}{6\pi} \int_0^{\frac{\pi}{2}} \frac{\sin(\theta) \cos(\theta)} {d^3(\theta)}  {\rm d}\theta  
\end{align}
Here $\theta$ denotes the zenith angle with the $-{\bf e}_z$ direction from the particle center and $d(\theta)$ the distance of the particle to the front on this direction (see Fig.~\ref{SketchParticle}).

The determination of $F_{\rm l}$ and $F_{\rm T}$ and thus of the features of the critical state at the trapping transition, requires to derive $d(\theta)$ from the Gibbs-Thomson relation (\ref{GT}). When surface tension is neglected, this cannot be handled analytically and requires numerical determinations~\cite{Rempel2001,Park2006}. 

Unfortunately, these numerical studies report a number of information on the velocity $V_p^c$ and the minimal distance $d_0^c$ to the front at the critical state, but do not report on the mean thermomolecular force. 
However, within some assumption regarding the geometry of the channel separating particle and front, we may obtain analytical relationships that provide a link between $V_p^c$, $d_0^c$ and $F_{\rm T}$ or $\bar{P_{\rm T}}$.

We consider below the Rempel and Worster assumption following which the channel between the particle and the front involves a constant width $d_0$
up to an angle $\theta_0$ beyond which it suddenly dramatically increases. This corresponds to a film climbing along the sphere and relaxing to a plane at a turning point, i.e to a concave-convex front shape around the particle (see Figure~\ref{SketchParticle}). This configuration echoes the change of front shape induced by the ratio of thermal conductivities $k_{\rm p}/k_{\rm m}$  between the particle and the medium. In particular, numerical studies by Park, Golovin and Davis~\cite{Park2006} have shown that, for our parameters $\gamma = T_{\rm M} \sigma/ L_v G R^2 = 2.68$ and $k_{\rm p}/k_{\rm m}=0.205$, the front shape assumes a concave-convex shape 
which fits the assumption of the Rempel-Worster model.

Within this constant film thickness assumption, $d(\theta)=d_0$ for $\theta<\theta_0$ and 
$d(\theta)=\infty$ beyond, the integrals (\ref{Fl}) and (\ref{FT}) can then be determined. Anticipating a small value of $\theta_0$, they yield, for $\theta_0 \ll \pi/2$ : 
\begin{align}
\label{PT,RW}
\bar{P_{\rm T}} &= \frac{\mathcal{A}}{6 \pi} \, \frac{ \theta_0^2 }{d_0^3} \\
\label{Pl,RW}
\bar{P_{\rm l}} &= - \frac{3}{2} \, \mu R^2  V_p \, \frac{\theta_0^4}{d_0^3} 
\end{align}
At mechanical equilibrium, both yield:
\begin{equation}
\label{thetaV}
\theta_0^2 V_p = \frac{\mathcal{A}}{6 \pi} \,  \frac{2}{3 \mu R^2} 
\end{equation}
On the other hand, the numerical determination of Rempel and Worster \cite{Rempel2001} yields for the film thickness $d_0^c$ and the velocity $V_p^c=V_p(d_0^c)$ at the critical state of the trapping transition :
\begin{align}
\label{Vc,RW}
V_p^c &= v_{\rm p}^c \left ( \frac{\mathcal{A}}{6\pi} \right )^{2/3} \! \frac{\sigma^{1/3}}{6\mu R^{4/3}}\\
\label{d0,RW}
d_0^c &= \delta_0^c \left ( \frac{\mathcal{A}}{6 \pi} \frac{R}{\sigma} \right)^{1/3}.
\end{align}
with, for the present parameters of the suspension ($\gamma=2.68$), $v_p^c \approx0,95$ and $\delta_0^c\approx 0.95$.

Our goal is now to obtain from this an evaluation of the mean thermomolecular pressure $\bar{P_{\rm T}} $ at the critical state.
For this, we use the relation (\ref{thetaV}) to determine the critical value $\theta_0^c$ of  $\theta_0$ which, together with $d_0^c$,
 then provides from (\ref{PT,RW}) the value of $\bar{P_{\rm T}}$ in the critical state. We obtain:
\begin{align}
\label{theta0,RW}
{\theta_0^c}^2 &= \frac{4}{v_{\rm p}^c} \left ( \frac{\mathcal{A}}{6\pi} \right )^{1/3} \! (\sigma R^2)^{-1/3}\\
\label{PT,RW,uc}
\bar{P_{\rm T}} &= \frac{4}{v_{\rm p}^c {\delta_0^c}^3} \left ( \frac{\mathcal{A}}{6 \pi} \right )^{1/3} \! \sigma^{2/3} R^{-5/3}
\end{align}
Relation (\ref{PT,RW,uc}) may be used here to provide an order of magnitude of the mean thermomolecular pressure exerted on a single entering particle in conditions similar to our experiment. It yields $\bar{P_{\rm T}} = 208$~Pa with, from relations (\ref{Vc,RW}) (\ref{d0,RW}), $V_p^c= 12.29$\mups and $d_0^c = 3.1$~nm. The latter value yields a thermomolecular pressure of $P_{\rm T0} = 22.5$~kPa at the base of the particle and up to the angular position $\theta_0$ in the present constant thickness model. The large relative difference between $\bar{P_{\rm T}}$ and $P_{\rm T0}$ then means that $P_{\rm T0}$ is indeed applied over a small angle $\theta_0$. This is confirmed from relation (\ref{theta0,RW}): $\theta_0^c = 0.096$~rad in the critical state.

\section{Discussion}
\label{Discussion}

The experimental results as well as their comparison with the Rempel-Worster model~\cite{Rempel2001}, hereafter called RW model, yield both qualitative and quantitative outcomes.

We first notice that the mean pressure $\bar{P}_\mu$ on particles in the compacted layer is lower than the mean thermomolecular pressure $\bar{P_{\rm T}} \approx 1$ ~kPa. It is thus significantly smaller than the pressure threshold of about $10$~kPa~\cite{Meijer1978,Goehring2010} above which a particle layer would turn aggregated
~\cite{Anderson2012,Anderson2014}. Accordingly, the particle layer, although compact, may be considered as dispersed, i.e. composed of freely moving particles, as confirmed by direct observation (see movie in supplementary material).
This supports the modeling of trapping based on individual particles rather than on a cohesive layer.

On a qualitative level, the affine relation of $h$ with $1/|U|$ supports the relevance, at least at the dominant order, of the mechanical model involving a viscous force proportional to $hU$, a lubrication force proportional to $U$ and a constant thermomolecular force. 

On a quantitative level, the measured mean thermomolecular pressure $\bar{P_{\rm T}} \approx 1$~kPa is found five times greater than in the constant film thickness RW model, $\bar{P_{\rm T}}=0.2$~kPa. However, both appear much smaller than the thermomolecular pressure $P_{\rm T0}=22.5$~kPa exerted in this model at the base of the particle and in the whole constant thickness film. This large difference means that our experiment supports the main feature of the particle-front interaction worked out in the RW model: a very large thermomolecular pressure exerted on a very small part of the particle surface.

However, beyond this agreement, the factor five difference between the magnitudes of the mean thermomolecular pressure deduced from experiment or predicted by the RW model questions the fine modeling of the repelling/trapping mechanism. In particular, an important difference between 
experiment and model is that our experiment addresses a particle layer whereas the RW model concerns a single isolated particle.
This yields in the experiment an additional force on an entering particle that is not involved in the RW model: the viscous force generated on the whole particle layer. We investigate below 
its implication on critical state features at the trapping transition as well as on the other possible origins of the factor five difference regarding the mean thermomolecular pressure.

We thus address the main differences between the RW model and the experiment so as to interpret their quantitative discrepancies. 
Some are fundamental: the implication of a particle layer on the film thickness.
Some refer to a difference of parameters: the Hamaker constant (or $\epsilon_S$ in~\cite{Rempel2001}) and the difference of thermal conductivity between particles and water. We review them below and determine the sense and the order of magnitude of their implication on $\bar{P_{\rm T}}$.

\subsection{Particle layer effect on film thickness}
\label{ParticleLayerEffect}

We address here the effect of the particle layer on the characteristics of the critical state of the trapping transition. On a particle adjacent to the front, the mechanical balance of forces yields, from section \ref{MechanicalBalance}, a resultant mean pressure :
\begin{equation}
\label{PT,U}
\bar{P} = - \frac{\mu}{\phi k} h |U| - f  V_p + \bar{P_{\rm T}}
\end{equation}

Following the Gibbs-Thomson relation (\ref{GT}), the geometry of the channel separating the entering particle and the front may be parametrized by its thickness $d_0$ at the base of the particle. 
Accordingly, the integrals involved in the  lubrication and thermomolecular forces (\ref{Fl})  (\ref{FT}) appear as functions of $d_0$ : $f\equiv f(d_0)$, $\bar{P_{\rm T}} \equiv \bar{P_{\rm T}}(d_0)$.

The equilibrium states, $\bar{P}=0$, may refer either to repelling or to trapping. The transition between them corresponds to a critical state. It may be reached on a single particle by increasing the solidification velocity $V$ or, in the present experiment where many particles are involved, by increasing the particle layer $h$.
The criterion for the transition may be obtained by considering the implications of fluctuations of the film thickness parameter $d_0$ : a reduction of $d_0$ will enforce trapping if the resultant ${\rm d}\bar{P}$ is negative ; it will enforce repelling if ${\rm d}\bar{P}$ is positive ; at the transition between these regimes, ${\rm d}\bar{P}$ thus vanishes, so that $V_p=\bar{P_{\rm T}}'/f'$, where the prime denotes the first derivative with respect to $d_0$.

This relation shows that $d_0$ and $V_p$ are linked on the critical states for trapping. It thus applies to the steady states of the particle layer on which the layer thickness $h$ and $V_p=|U|$ are related one to the other (see Figure~\ref{G,h,1-U}).
Altogether, this shows $h$ and $d_0$ are linked together : the additional force exerted on an entering particle by the compacted layer modifies the thickness of the film that separates it from the front. As a consequence, the mean thermomolecular pressure $\bar{P_{\rm T}}$ on it should raise accordingly.

To evaluate the order of magnitude of this multiple particle effect on $d_0$ and $\bar{P_{\rm T}}$, we may consider the analytical determinations of $P_T$ and $f$ derived when surface tension is neglected \cite{Rempel1999,Park2006}. They correspond to polynomial expressions \footnote{$\bar{P_{\rm T}} \propto(\delta_0^{-2}+\delta_0^{-6})$ ; $f \propto (\delta_0^{-1}+ 4/5 \delta_0^{-5}+ 1/3 \delta_0^{-9})$, where $\delta_0$ stands for the non-dimensional film thickness $d_0/l$ with $l^4=\mathcal{A}/6\pi \; T_M/L_v$.} from which $V_p(d_0)=\bar{P_{\rm T}}'/f'$ may be deduced. 
Denoting, as in section \ref{SingleParticleCriticalState}, the critical velocity for single particle trapping by $V_p^c$ and its critical film thickness by $d_0^c$, it then appears that, down to $V_p^c/10$, the relationship between $V_p$ and $d_0$ may be approached by $V_p= V_p^c [1- 3/2 \; (1-d_0/d_0^c)]$.
In our experiment, as a vanishing layer thickness was obtained for $V\approx 20$\mups{} at $\phi_0=0.2$, i.e for $U\approx 14$\mups{}, one may estimate $V_p^c$ at about $15$\mups{}. In comparison, the domain on which the determination of $\bar{P_{\rm T}}$ has been obtained stands at velocities around $V_p^c/5$ for which $d_0$ is decreased by a half, and the mean thermomolecular pressure $\bar{P_{\rm T}}$ raised by a factor $40$.

Although this estimation stands for a vanishing surface tension, it indicates that, in our experiment where a particle layer is involved, the measured mean thermomolecular pressure $\bar{P_{\rm T}}$ on an entering particle is likely raised by a factor of several tens in comparison to its value on a single isolated particle.

\subsection{Hamaker constant}

Models of particle trapping are based on a numerical determination of the front shape from the Gibbs-Thomson relation (\ref{GT}). Three kinds of effects then affect the front position: (i) the effect of the thermal gradient given the front undercooling; (ii) capillary effects on undercooling; and (iii) thermomolecular pressure on undercooling, the solutal effect being neglected. These effects are respectively conveyed by the three variables $G$, $\sigma$ and $\mathcal{A}$ or equivalently by two non-dimensional parameters, $\gamma=\sigma T_{\rm M}/Lv G R^2$ and either 
$\epsilon=(\mathcal{A} T_{\rm M}/6 \pi L_v G R^4)^{1/4}$~\cite{Rempel1999,Park2006} or $\epsilon_{\rm S}=(\mathcal{A}/6 \pi \sigma R^2)^{1/3}$ \cite{Rempel2001}
 depending on studies.
 
 However, the numerical solutions of these models are provided in terms of graphs parametrized by only one of these parameters, $\gamma$~\cite{Rempel1999,Rempel2001,Park2006}. In particular, the RW model~\cite{Rempel2001} on which the analysis of Section~\ref{SingleParticleCriticalState} relies, provides evolution of non-dimensional velocity and film thickness at $\epsilon_{\rm S}=10^{-3}$ whereas our experiments are conducted at $\epsilon_{\rm S}=2.9 \times 10^{-3}$. As $\epsilon_{\rm S}$ scales as $\mathcal{A}^{1/3}$, this may be interpreted as an Hamaker constant $2.9^3 \approx 25$ larger in experiment than in simulations. Equivalently, this means that the net thermomolecular pressure $\bar{P_{\rm T}}$ that should be expected from a more accurate modeling should be increased accordingly.

In particular, assuming that the non-dimensional variables $\delta_0^c$ and $v_{\rm p}^c$ are constant under this variation of $\mathcal{A}$, relation (\ref{PT,RW,uc}) states that $\bar{P_{\rm T}}$ scales as $\mathcal{A}^{1/3}$, i.e. as $\epsilon_{\rm S}$, and should therefore be enhanced by a factor $25^{1/3} \approx 2.9$.

\subsection{Thermal conductivity differences}

Particles are made of polystyrene of a lower thermal conductivity $k_{\rm p}=0.115$ than that of water $k_{\rm m}=0.561$.
The resulting modification of the thermal field in their vicinity tends to reject the solidification front away from the particle and form a bulge below the particle. This would dramatically increase the film thickness $d$ and decrease $\bar{P_{\rm T}}$. However, this tendency is partly balanced by capillarity which acts so as to flatten the front.
For our experimental parameters ($\gamma=2.68$, $k_{\rm p}/k_{\rm m}=0.205$), the outcome appears as a concave-convex interface which follows the particle shape up to a turning point where it relaxes towards the planar front~\cite{Park2006}. This provides a film thickness somewhat constant over an angular sector below the particle and beyond which it sharply increases (see Figure~\ref{SketchParticle}), in agreement with the geometric features of the modeling considered by Rempel and Worster~\cite{Rempel2001} and used in section~\ref{SingleParticleCriticalState}.

The implication of a thermal conductivity mismatch has been addressed by Park, Golovin and Davis~\cite{Park2006} on both the velocity $V_p^c$ and the film thickness $d_{0}^c$ but not the mean thermomolecular force. However, it may be deduced from the former variables within the assumption of a constant film thickness over an angular sector $\theta_0$, using the analysis provided in section~\ref{SingleParticleCriticalState}. Following the numerical results of reference~\cite{Park2006}, the change at $\gamma=1$ of the ratio $k_{\rm p}/k_{\rm m}$ from $1$, as implicitly considered up to now, to $0.1$ (resp. $10$) yields an increase of $V_p^c$ by a factor $7.8$ (resp. $4.3 \times 10^{-2}$) and of $d_{0}^c$ by a factor $1.7$ (resp. $0.5$). Accordingly, for the experimental parameters $k_{\rm p}/k_{\rm m}=0.205$, $\gamma \approx 1$, one may estimate, by a power law extrapolation, the corresponding increase factors for $V_c$ and $d_{0}^c$ as $\alpha=3.94$ and $\beta=1.45$ respectively. Following relation (\ref{thetaV}), this implies a reduction of $\theta_0^2$ by the factor $\alpha$ and thus, from (\ref{PT,RW}), a reduction of $\bar{P_{\rm T}}$ by a factor $\alpha \beta^3$, i.e. about $12$.

\vskip\baselineskip

\subsection{Synthesis}

The value $\bar{P_{\rm T}}=208$~Pa deduced in section~\ref{SingleParticleCriticalState} from an extrapolation based on the RW model for a single particle needs thus to be corrected by definite factors for several reasons :
 i) a fundamental reason relative to the particle layer interplay with the critical state features (factor $\approx 40$) ;
 ii) parameter shifts relative to the parameter $\epsilon_S$ (or to the Hamaker constant) in~\cite{Rempel2001} (factor $3$) and to thermal conductivity differences between particle and water (factor $12^{-1}$). Altogether, this yields a reevaluation of the $\bar{P_{\rm T}}$ estimate by a factor about $40\times 3 / 12$, i.e. about $10$.

This moderately large factor illustrates that several agonist and antagonist phenomena take part in the design of the film which separates the particle and the front and, thus, in the net thermomolecular pressure applied on the particle. The correction factor, of order $10$, agrees with the observed difference between the experimental measure of $\bar{P_{\rm T}}$ ($\approx$~1.0 kPa) and its primary estimation by modeling ($\approx 0.2$~kPa). On the other hand, it is weak enough to maintain a large gap between such estimations and that of the thermomolecular pressure at the base of an entering particle ($\approx 20$~kPa), thereby corroborating that a small part of the particle surface supports the main thermomolecular pressure.

These results support the thickness of the particle layer as a relevant indicator of the physical conditions experienced by the entering particles. Interestingly, we notice that this variable involves a macroscopic magnitude as well as a statistical nature implied by the large number of particles hold in the layer or entering to the front.
Altogether, it thus provides a macroscopic probe of the averaged trapping mechanism on a solidification front.

\section{Conclusion}
\label{Conclusion}

This study addressed the solidification of suspensions below the trapping velocity of single isolated particles. Particles were found to accumulate in a particle layer whose thickness growth rate reveals its close-packing. This implies multiple particle interactions both in the particle layer and with the front.

At any velocity, particle trapping occurred for a sufficiently large thickness of the layer. It then compensated particle accumulation and made the layer growth stop.  We focussed attention on this steady-state layer thickness to provide insights into the multiple particle interactions and the resulting particle trapping mechanism.

Two kinds of multiple particle effects have been identified.
In the compacted particle layer, viscous friction on the particle matrix and the associated fluid pressure drop induce an additional force on particles which favors trapping. On the other hand, at the level of a particle entering the front, this additional force leads a reduction of the film that separates particle and front with a corresponding raise of the repelling thermomolecular force exerted by front on particles.

A quantitative study was provided by a mechanical model on particle adjacent to the front which includes the additional force induced by the particle layer. 
It provided a mean thermomolecular pressure on an entering particle that is low compared to that exerted at its base and large compared to that proposed by models for a single isolated particle. 
The former feature means that a small part of the particle surface undergoes a large repelling from the front, the film between them thus enlarging quickly beyond.
The latter statement was explained by the reduction of film thickness in response to the additional force induced by multiple particle effects.

Altogether, the present study thus evidenced the effects of multiple particle interactions on particle trapping by solidification fronts.
In particular, it clarified the formation of a compacted layer of particles ahead of the solidification front and pointed out its role and its effect on the trapping mechanism.
For the present particle features and solidification conditions, the solidification front was mainly planar and the particle layer disperse ; for other particle diameters or solidification velocities, some patterning could occur in either the particle layer, the front or the frozen phase and the layer could turn aggregated. As all these events largely result from multiple particles interaction with the front, the findings reported here will thus provide a relevant basis for understanding their origin and their inner mechanisms.

\section*{ASSOCIATED CONTENT}
\textbf{ Supporting Information}\\
A supporting movie, along with its short description. This material is available free of charge via the Internet at http://pubs.acs.org.

\section*{Acknowledgments}
The research leading to these results has been supported by the European Research Council under the European Union's Seventh Framework Program (FP7/2007-2013)/ERC grant agreement 278004.
    
\bibliography{ReferenceSolidificationSuspension}

\begin{thebibliography}{49}%
\makeatletter
\providecommand \@ifxundefined [1]{%
 \@ifx{#1\undefined}
}%
\providecommand \@ifnum [1]{%
 \ifnum #1\expandafter \@firstoftwo
 \else \expandafter \@secondoftwo
 \fi
}%
\providecommand \@ifx [1]{%
 \ifx #1\expandafter \@firstoftwo
 \else \expandafter \@secondoftwo
 \fi
}%
\providecommand \natexlab [1]{#1}%
\providecommand \enquote  [1]{``#1''}%
\providecommand \bibnamefont  [1]{#1}%
\providecommand \bibfnamefont [1]{#1}%
\providecommand \citenamefont [1]{#1}%
\providecommand \href@noop [0]{\@secondoftwo}%
\providecommand \href [0]{\begingroup \@sanitize@url \@href}%
\providecommand \@href[1]{\@@startlink{#1}\@@href}%
\providecommand \@@href[1]{\endgroup#1\@@endlink}%
\providecommand \@sanitize@url [0]{\catcode `\\12\catcode `\$12\catcode
  `\&12\catcode `\#12\catcode `\^12\catcode `\_12\catcode `\%12\relax}%
\providecommand \@@startlink[1]{}%
\providecommand \@@endlink[0]{}%
\providecommand \url  [0]{\begingroup\@sanitize@url \@url }%
\providecommand \@url [1]{\endgroup\@href {#1}{\urlprefix }}%
\providecommand \urlprefix  [0]{URL }%
\providecommand \Eprint [0]{\href }%
\providecommand \doibase [0]{http://dx.doi.org/}%
\providecommand \selectlanguage [0]{\@gobble}%
\providecommand \bibinfo  [0]{\@secondoftwo}%
\providecommand \bibfield  [0]{\@secondoftwo}%
\providecommand \translation [1]{[#1]}%
\providecommand \BibitemOpen [0]{}%
\providecommand \bibitemStop [0]{}%
\providecommand \bibitemNoStop [0]{.\EOS\space}%
\providecommand \EOS [0]{\spacefactor3000\relax}%
\providecommand \BibitemShut  [1]{\csname bibitem#1\endcsname}%
\let\auto@bib@innerbib\@empty
\bibitem [{\citenamefont {Corte}(1962)}]{Corte1962}%
  \BibitemOpen
  \bibfield  {author} {\bibinfo {author} {\bibfnamefont {A.~E.}\ \bibnamefont
  {Corte}},\ }\href {\doibase 10.1029/JZ067i003p01085} {\bibfield  {journal}
  {\bibinfo  {journal} {Journal of Geophysical Research}\ }\textbf {\bibinfo
  {volume} {67}},\ \bibinfo {pages} {1085} (\bibinfo {year}
  {1962})}\BibitemShut {NoStop}%
\bibitem [{\citenamefont {Zhu}\ \emph {et~al.}(2000)\citenamefont {Zhu},
  \citenamefont {Vilches}, \citenamefont {Dash}, \citenamefont {Sing},\ and\
  \citenamefont {Wettlaufer}}]{Zhu2000}%
  \BibitemOpen
  \bibfield  {author} {\bibinfo {author} {\bibfnamefont {D.}~\bibnamefont
  {Zhu}}, \bibinfo {author} {\bibfnamefont {O.}~\bibnamefont {Vilches}},
  \bibinfo {author} {\bibfnamefont {J.}~\bibnamefont {Dash}}, \bibinfo {author}
  {\bibfnamefont {B.}~\bibnamefont {Sing}}, \ and\ \bibinfo {author}
  {\bibfnamefont {J.}~\bibnamefont {Wettlaufer}},\ }\href@noop {} {\bibfield
  {journal} {\bibinfo  {journal} {Physical Review Letters}\ }\textbf {\bibinfo
  {volume} {85}},\ \bibinfo {pages} {4908} (\bibinfo {year}
  {2000})}\BibitemShut {NoStop}%
\bibitem [{\citenamefont {Rempel}\ \emph {et~al.}(2004)\citenamefont {Rempel},
  \citenamefont {Wettlaufer},\ and\ \citenamefont {Worster}}]{Rempel2004}%
  \BibitemOpen
  \bibfield  {author} {\bibinfo {author} {\bibfnamefont {A.~W.}\ \bibnamefont
  {Rempel}}, \bibinfo {author} {\bibfnamefont {J.~S.}\ \bibnamefont
  {Wettlaufer}}, \ and\ \bibinfo {author} {\bibfnamefont {M.~G.}\ \bibnamefont
  {Worster}},\ }\href {\doibase 10.1017/S0022112003006761} {\bibfield
  {journal} {\bibinfo  {journal} {Journal of Fluid Mechanics}\ }\textbf
  {\bibinfo {volume} {498}},\ \bibinfo {pages} {227} (\bibinfo {year}
  {2004})}\BibitemShut {NoStop}%
\bibitem [{\citenamefont {Dash}\ \emph {et~al.}(2006)\citenamefont {Dash},
  \citenamefont {Rempel},\ and\ \citenamefont {Wettlaufer}}]{Dash2006}%
  \BibitemOpen
  \bibfield  {author} {\bibinfo {author} {\bibfnamefont {J.}~\bibnamefont
  {Dash}}, \bibinfo {author} {\bibfnamefont {A.~W.}\ \bibnamefont {Rempel}}, \
  and\ \bibinfo {author} {\bibfnamefont {J.~S.}\ \bibnamefont {Wettlaufer}},\
  }\href {\doibase 10.1103/RevModPhys.78.695} {\bibfield  {journal} {\bibinfo
  {journal} {Reviews of Modern Physics}\ }\textbf {\bibinfo {volume} {78}},\
  \bibinfo {pages} {695} (\bibinfo {year} {2006})}\BibitemShut {NoStop}%
\bibitem [{\citenamefont {Peppin}\ and\ \citenamefont
  {Style}(2013)}]{Peppin2013}%
  \BibitemOpen
  \bibfield  {author} {\bibinfo {author} {\bibfnamefont {S.~S.~L.}\
  \bibnamefont {Peppin}}\ and\ \bibinfo {author} {\bibfnamefont {R.~W.}\
  \bibnamefont {Style}},\ }\href {\doibase 10.2136/vzj2012.0049} {\bibfield
  {journal} {\bibinfo  {journal} {Vadose Zone Journal}\ }\textbf {\bibinfo
  {volume} {12}} (\bibinfo {year} {2013}),\ 10.2136/vzj2012.0049}\BibitemShut
  {NoStop}%
\bibitem [{\citenamefont {Saruya}\ \emph {et~al.}(2013)\citenamefont {Saruya},
  \citenamefont {Kurita},\ and\ \citenamefont {Rempel}}]{Saruya2013}%
  \BibitemOpen
  \bibfield  {author} {\bibinfo {author} {\bibfnamefont {T.}~\bibnamefont
  {Saruya}}, \bibinfo {author} {\bibfnamefont {K.}~\bibnamefont {Kurita}}, \
  and\ \bibinfo {author} {\bibfnamefont {A.~W.}\ \bibnamefont {Rempel}},\
  }\href {\doibase 10.1103/PhysRevE.87.032404} {\bibfield  {journal} {\bibinfo
  {journal} {Physical Review E}\ }\textbf {\bibinfo {volume} {87}},\ \bibinfo
  {pages} {032404} (\bibinfo {year} {2013})}\BibitemShut {NoStop}%
\bibitem [{\citenamefont {Anderson}\ and\ \citenamefont {{Grae
  Worster}}(2014)}]{Anderson2014}%
  \BibitemOpen
  \bibfield  {author} {\bibinfo {author} {\bibfnamefont {A.~M.}\ \bibnamefont
  {Anderson}}\ and\ \bibinfo {author} {\bibfnamefont {M.}~\bibnamefont {{Grae
  Worster}}},\ }\href {\doibase 10.1017/jfm.2014.500} {\bibfield  {journal}
  {\bibinfo  {journal} {Journal of Fluid Mechanics}\ }\textbf {\bibinfo
  {volume} {758}},\ \bibinfo {pages} {786} (\bibinfo {year}
  {2014})}\BibitemShut {NoStop}%
\bibitem [{\citenamefont {Saruya}\ \emph {et~al.}(2014)\citenamefont {Saruya},
  \citenamefont {Rempel},\ and\ \citenamefont {Kurita}}]{Saruya2014}%
  \BibitemOpen
  \bibfield  {author} {\bibinfo {author} {\bibfnamefont {T.}~\bibnamefont
  {Saruya}}, \bibinfo {author} {\bibfnamefont {A.~W.}\ \bibnamefont {Rempel}},
  \ and\ \bibinfo {author} {\bibfnamefont {K.}~\bibnamefont {Kurita}},\ }\href
  {\doibase 10.1021/jp505366y} {\bibfield  {journal} {\bibinfo  {journal} {The
  Journal of Physical Chemistry B}\ }\textbf {\bibinfo {volume} {118}},\
  \bibinfo {pages} {13420} (\bibinfo {year} {2014})}\BibitemShut {NoStop}%
\bibitem [{\citenamefont {Bronstein}\ \emph {et~al.}(1981)\citenamefont
  {Bronstein}, \citenamefont {Itkin},\ and\ \citenamefont
  {Ishkov}}]{Bronstein1981}%
  \BibitemOpen
  \bibfield  {author} {\bibinfo {author} {\bibfnamefont {V.}~\bibnamefont
  {Bronstein}}, \bibinfo {author} {\bibfnamefont {Y.}~\bibnamefont {Itkin}}, \
  and\ \bibinfo {author} {\bibfnamefont {G.}~\bibnamefont {Ishkov}},\
  }\href@noop {} {\bibfield  {journal} {\bibinfo  {journal} {Journal of Crystal
  Growth}\ }\textbf {\bibinfo {volume} {52}},\ \bibinfo {pages} {345} (\bibinfo
  {year} {1981})}\BibitemShut {NoStop}%
\bibitem [{\citenamefont {K{\"{o}}rber}(1988)}]{Korber1988}%
  \BibitemOpen
  \bibfield  {author} {\bibinfo {author} {\bibfnamefont {C.}~\bibnamefont
  {K{\"{o}}rber}},\ }\href {\doibase 10.1017/S0033583500004303} {\bibfield
  {journal} {\bibinfo  {journal} {Quarterly reviews of biophysics}\ }\textbf
  {\bibinfo {volume} {21}},\ \bibinfo {pages} {229} (\bibinfo {year}
  {1988})}\BibitemShut {NoStop}%
\bibitem [{\citenamefont {Velez-Ruiz}\ and\ \citenamefont
  {Rahman}(1981)}]{Velez-Ruiz2007}%
  \BibitemOpen
  \bibfield  {author} {\bibinfo {author} {\bibfnamefont {J.}~\bibnamefont
  {Velez-Ruiz}}\ and\ \bibinfo {author} {\bibfnamefont {M.}~\bibnamefont
  {Rahman}},\ }\bibfield  {booktitle} {\emph {\bibinfo {booktitle} {Handbook of
  food preservation}},\ }\href@noop {} {\ ,\ \bibinfo {pages} {345} (\bibinfo
  {year} {1981})}\BibitemShut {NoStop}%
\bibitem [{\citenamefont {Stefanescu}\ \emph {et~al.}(1988)\citenamefont
  {Stefanescu}, \citenamefont {Dhindaw}, \citenamefont {Kacar},\ and\
  \citenamefont {Moitra}}]{Stefanescu1988}%
  \BibitemOpen
  \bibfield  {author} {\bibinfo {author} {\bibfnamefont {D.~M.}\ \bibnamefont
  {Stefanescu}}, \bibinfo {author} {\bibfnamefont {B.~K.}\ \bibnamefont
  {Dhindaw}}, \bibinfo {author} {\bibfnamefont {S.~A.}\ \bibnamefont {Kacar}},
  \ and\ \bibinfo {author} {\bibfnamefont {A.}~\bibnamefont {Moitra}},\ }\href
  {\doibase 10.1007/BF02645819} {\bibfield  {journal} {\bibinfo  {journal}
  {Metallurgical Transactions A}\ }\textbf {\bibinfo {volume} {19}},\ \bibinfo
  {pages} {2847} (\bibinfo {year} {1988})}\BibitemShut {NoStop}%
\bibitem [{\citenamefont {Asthana}\ and\ \citenamefont
  {Tewari}(1993)}]{Asthana1993}%
  \BibitemOpen
  \bibfield  {author} {\bibinfo {author} {\bibfnamefont {R.}~\bibnamefont
  {Asthana}}\ and\ \bibinfo {author} {\bibfnamefont {S.~N.}\ \bibnamefont
  {Tewari}},\ }\href {\doibase 10.1007/BF00367810} {\bibfield  {journal}
  {\bibinfo  {journal} {Journal of Materials Science}\ }\textbf {\bibinfo
  {volume} {28}},\ \bibinfo {pages} {5414} (\bibinfo {year}
  {1993})}\BibitemShut {NoStop}%
\bibitem [{\citenamefont {Liu}\ \emph {et~al.}(2008)\citenamefont {Liu},
  \citenamefont {Nakano},\ and\ \citenamefont {Kakimoto}}]{Liu2008}%
  \BibitemOpen
  \bibfield  {author} {\bibinfo {author} {\bibfnamefont {L.}~\bibnamefont
  {Liu}}, \bibinfo {author} {\bibfnamefont {S.}~\bibnamefont {Nakano}}, \ and\
  \bibinfo {author} {\bibfnamefont {K.}~\bibnamefont {Kakimoto}},\ }\href
  {\doibase 10.1016/j.jcrysgro.2007.11.165} {\bibfield  {journal} {\bibinfo
  {journal} {Journal of Crystal Growth}\ }\textbf {\bibinfo {volume} {310}},\
  \bibinfo {pages} {2192} (\bibinfo {year} {2008})}\BibitemShut {NoStop}%
\bibitem [{\citenamefont {Deville}\ \emph {et~al.}(2007)\citenamefont
  {Deville}, \citenamefont {Saiz},\ and\ \citenamefont
  {Tomsia}}]{Deville2007a}%
  \BibitemOpen
  \bibfield  {author} {\bibinfo {author} {\bibfnamefont {S.}~\bibnamefont
  {Deville}}, \bibinfo {author} {\bibfnamefont {E.}~\bibnamefont {Saiz}}, \
  and\ \bibinfo {author} {\bibfnamefont {A.~P.}\ \bibnamefont {Tomsia}},\
  }\href {\doibase 10.1016/j.actamat.2006.11.003} {\bibfield  {journal}
  {\bibinfo  {journal} {Acta Materialia}\ }\textbf {\bibinfo {volume} {55}},\
  \bibinfo {pages} {1965} (\bibinfo {year} {2007})}\BibitemShut {NoStop}%
\bibitem [{\citenamefont {Chernov}\ \emph {et~al.}(1976)\citenamefont
  {Chernov}, \citenamefont {Temkin},\ and\ \citenamefont
  {Mel'nikova}}]{Chernov1976}%
  \BibitemOpen
  \bibfield  {author} {\bibinfo {author} {\bibfnamefont {A.~A.}\ \bibnamefont
  {Chernov}}, \bibinfo {author} {\bibfnamefont {D.~E.}\ \bibnamefont {Temkin}},
  \ and\ \bibinfo {author} {\bibfnamefont {A.~M.}\ \bibnamefont {Mel'nikova}},\
  }\href@noop {} {\bibfield  {journal} {\bibinfo  {journal} {Soviet Physics and
  Crystallography}\ }\textbf {\bibinfo {volume} {21}},\ \bibinfo {pages} {369}
  (\bibinfo {year} {1976})}\BibitemShut {NoStop}%
\bibitem [{\citenamefont {Wettlaufer}\ and\ \citenamefont
  {Worster}(2006)}]{Wettlaufer2006}%
  \BibitemOpen
  \bibfield  {author} {\bibinfo {author} {\bibfnamefont {J.~S.}\ \bibnamefont
  {Wettlaufer}}\ and\ \bibinfo {author} {\bibfnamefont {M.~G.}\ \bibnamefont
  {Worster}},\ }\href {\doibase 10.1146/annurev.fluid.37.061903.175758}
  {\bibfield  {journal} {\bibinfo  {journal} {Annual Review of Fluid
  Mechanics}\ }\textbf {\bibinfo {volume} {38}},\ \bibinfo {pages} {427}
  (\bibinfo {year} {2006})}\BibitemShut {NoStop}%
\bibitem [{\citenamefont {Zhang}\ \emph {et~al.}(2005)\citenamefont {Zhang},
  \citenamefont {Hussain}, \citenamefont {Brust}, \citenamefont {Butler},
  \citenamefont {Rannard},\ and\ \citenamefont {Cooper}}]{Zhang2005}%
  \BibitemOpen
  \bibfield  {author} {\bibinfo {author} {\bibfnamefont {H.}~\bibnamefont
  {Zhang}}, \bibinfo {author} {\bibfnamefont {I.}~\bibnamefont {Hussain}},
  \bibinfo {author} {\bibfnamefont {M.}~\bibnamefont {Brust}}, \bibinfo
  {author} {\bibfnamefont {M.~F.}\ \bibnamefont {Butler}}, \bibinfo {author}
  {\bibfnamefont {S.~P.}\ \bibnamefont {Rannard}}, \ and\ \bibinfo {author}
  {\bibfnamefont {A.~I.}\ \bibnamefont {Cooper}},\ }\href {\doibase
  10.1038/nmat1487} {\bibfield  {journal} {\bibinfo  {journal} {Nature
  Materials}\ }\textbf {\bibinfo {volume} {4}},\ \bibinfo {pages} {787}
  (\bibinfo {year} {2005})}\BibitemShut {NoStop}%
\bibitem [{\citenamefont {Youssef}\ \emph {et~al.}(2005)\citenamefont
  {Youssef}, \citenamefont {Dashwood},\ and\ \citenamefont
  {Lee}}]{Youssef2005}%
  \BibitemOpen
  \bibfield  {author} {\bibinfo {author} {\bibfnamefont {Y.}~\bibnamefont
  {Youssef}}, \bibinfo {author} {\bibfnamefont {R.}~\bibnamefont {Dashwood}}, \
  and\ \bibinfo {author} {\bibfnamefont {P.}~\bibnamefont {Lee}},\ }\href@noop
  {} {\bibfield  {journal} {\bibinfo  {journal} {Composites Part A - Applied
  Science and Manufacturing}\ }\textbf {\bibinfo {volume} {36}},\ \bibinfo
  {pages} {747} (\bibinfo {year} {2005})}\BibitemShut {NoStop}%
\bibitem [{\citenamefont {Deville}\ \emph {et~al.}(2006)\citenamefont
  {Deville}, \citenamefont {Saiz}, \citenamefont {Nalla},\ and\ \citenamefont
  {Tomsia}}]{Deville2006c}%
  \BibitemOpen
  \bibfield  {author} {\bibinfo {author} {\bibfnamefont {S.}~\bibnamefont
  {Deville}}, \bibinfo {author} {\bibfnamefont {E.}~\bibnamefont {Saiz}},
  \bibinfo {author} {\bibfnamefont {R.~K.}\ \bibnamefont {Nalla}}, \ and\
  \bibinfo {author} {\bibfnamefont {A.~P.}\ \bibnamefont {Tomsia}},\ }\href
  {\doibase 10.1126/science.1120937} {\bibfield  {journal} {\bibinfo  {journal}
  {Science}\ }\textbf {\bibinfo {volume} {311}},\ \bibinfo {pages} {515 }
  (\bibinfo {year} {2006})}\BibitemShut {NoStop}%
\bibitem [{\citenamefont {Deville}\ \emph {et~al.}(2009)\citenamefont
  {Deville}, \citenamefont {Maire}, \citenamefont {Bernard-Granger},
  \citenamefont {Lasalle}, \citenamefont {Bogner}, \citenamefont {Gauthier},
  \citenamefont {Leloup},\ and\ \citenamefont {Guizard}}]{Deville2009}%
  \BibitemOpen
  \bibfield  {author} {\bibinfo {author} {\bibfnamefont {S.}~\bibnamefont
  {Deville}}, \bibinfo {author} {\bibfnamefont {E.}~\bibnamefont {Maire}},
  \bibinfo {author} {\bibfnamefont {G.}~\bibnamefont {Bernard-Granger}},
  \bibinfo {author} {\bibfnamefont {A.}~\bibnamefont {Lasalle}}, \bibinfo
  {author} {\bibfnamefont {A.}~\bibnamefont {Bogner}}, \bibinfo {author}
  {\bibfnamefont {C.}~\bibnamefont {Gauthier}}, \bibinfo {author}
  {\bibfnamefont {J.}~\bibnamefont {Leloup}}, \ and\ \bibinfo {author}
  {\bibfnamefont {C.}~\bibnamefont {Guizard}},\ }\href {\doibase
  10.1038/nmat2571} {\bibfield  {journal} {\bibinfo  {journal} {Nature
  Materials}\ }\textbf {\bibinfo {volume} {8}},\ \bibinfo {pages} {966}
  (\bibinfo {year} {2009})}\BibitemShut {NoStop}%
\bibitem [{\citenamefont {K{\"{o}}rber}\ \emph {et~al.}(1985)\citenamefont
  {K{\"{o}}rber}, \citenamefont {Rau},\ and\ \citenamefont
  {Others}}]{Korber1985}%
  \BibitemOpen
  \bibfield  {author} {\bibinfo {author} {\bibfnamefont {C.}~\bibnamefont
  {K{\"{o}}rber}}, \bibinfo {author} {\bibfnamefont {M.}~\bibnamefont {Rau}}, \
  and\ \bibinfo {author} {\bibnamefont {Others}},\ }\href
  {http://linkinghub.elsevier.com/retrieve/pii/0022024885902179} {\bibfield
  {journal} {\bibinfo  {journal} {Journal of crystal growth}\ }\textbf
  {\bibinfo {volume} {72}},\ \bibinfo {pages} {649} (\bibinfo {year}
  {1985})}\BibitemShut {NoStop}%
\bibitem [{\citenamefont {Ciss{\'{e}}}(1971)}]{Cisse1971}%
  \BibitemOpen
  \bibfield  {author} {\bibinfo {author} {\bibfnamefont {J.}~\bibnamefont
  {Ciss{\'{e}}}},\ }\href {\doibase 10.1016/0022-0248(71)90047-9} {\bibfield
  {journal} {\bibinfo  {journal} {Journal of Crystal Growth}\ }\textbf
  {\bibinfo {volume} {10}},\ \bibinfo {pages} {67} (\bibinfo {year}
  {1971})}\BibitemShut {NoStop}%
\bibitem [{\citenamefont {Zubkho}\ \emph {et~al.}(1973)\citenamefont {Zubkho},
  \citenamefont {Lobanov},\ and\ \citenamefont {Nikonova}}]{Zubkho1973}%
  \BibitemOpen
  \bibfield  {author} {\bibinfo {author} {\bibfnamefont {A.}~\bibnamefont
  {Zubkho}}, \bibinfo {author} {\bibfnamefont {V.}~\bibnamefont {Lobanov}}, \
  and\ \bibinfo {author} {\bibfnamefont {V.}~\bibnamefont {Nikonova}},\
  }\href@noop {} {\bibfield  {journal} {\bibinfo  {journal} {Soviet Physics and
  Crystallography}\ }\textbf {\bibinfo {volume} {18}},\ \bibinfo {pages} {239}
  (\bibinfo {year} {1973})}\BibitemShut {NoStop}%
\bibitem [{\citenamefont {Sen}\ \emph {et~al.}(1997)\citenamefont {Sen},
  \citenamefont {Kaukler}, \citenamefont {Curreri},\ and\ \citenamefont
  {Stefanescu}}]{Sen1997}%
  \BibitemOpen
  \bibfield  {author} {\bibinfo {author} {\bibfnamefont {S.}~\bibnamefont
  {Sen}}, \bibinfo {author} {\bibfnamefont {W.}~\bibnamefont {Kaukler}},
  \bibinfo {author} {\bibfnamefont {P.}~\bibnamefont {Curreri}}, \ and\
  \bibinfo {author} {\bibfnamefont {D.}~\bibnamefont {Stefanescu}},\
  }\href@noop {} {\bibfield  {journal} {\bibinfo  {journal} {Metallurgical
  Transactions A}\ }\textbf {\bibinfo {volume} {28}},\ \bibinfo {pages} {2129}
  (\bibinfo {year} {1997})}\BibitemShut {NoStop}%
\bibitem [{\citenamefont {Chernov}\ \emph {et~al.}(1977)\citenamefont
  {Chernov}, \citenamefont {Temkin},\ and\ \citenamefont
  {Mel'nikova}}]{Chernov1977}%
  \BibitemOpen
  \bibfield  {author} {\bibinfo {author} {\bibfnamefont {A.~A.}\ \bibnamefont
  {Chernov}}, \bibinfo {author} {\bibfnamefont {D.~E.}\ \bibnamefont {Temkin}},
  \ and\ \bibinfo {author} {\bibfnamefont {A.~M.}\ \bibnamefont {Mel'nikova}},\
  }\href@noop {} {\bibfield  {journal} {\bibinfo  {journal} {Sov. Phys.
  Crystallogr.}\ }\textbf {\bibinfo {volume} {22}},\ \bibinfo {pages} {656}
  (\bibinfo {year} {1977})}\BibitemShut {NoStop}%
\bibitem [{\citenamefont {Gilpin}(1980)}]{Gilpin1980}%
  \BibitemOpen
  \bibfield  {author} {\bibinfo {author} {\bibfnamefont {R.~R.}\ \bibnamefont
  {Gilpin}},\ }\href {\doibase 10.1029/WR016i005p00918} {\bibfield  {journal}
  {\bibinfo  {journal} {Water Resources Research}\ }\textbf {\bibinfo {volume}
  {16}},\ \bibinfo {pages} {918} (\bibinfo {year} {1980})}\BibitemShut
  {NoStop}%
\bibitem [{\citenamefont {Uhlmann}\ \emph {et~al.}(1964)\citenamefont
  {Uhlmann}, \citenamefont {Chalmers},\ and\ \citenamefont
  {Jackson}}]{Uhlmann1964}%
  \BibitemOpen
  \bibfield  {author} {\bibinfo {author} {\bibfnamefont {D.~R.}\ \bibnamefont
  {Uhlmann}}, \bibinfo {author} {\bibfnamefont {B.}~\bibnamefont {Chalmers}}, \
  and\ \bibinfo {author} {\bibfnamefont {K.~A.}\ \bibnamefont {Jackson}},\
  }\href {\doibase 10.1063/1.1713142} {\bibfield  {journal} {\bibinfo
  {journal} {Journal of Applied Physics}\ }\textbf {\bibinfo {volume} {35}},\
  \bibinfo {pages} {2986} (\bibinfo {year} {1964})}\BibitemShut {NoStop}%
\bibitem [{\citenamefont {Hadji}(1999)}]{Hadji1999}%
  \BibitemOpen
  \bibfield  {author} {\bibinfo {author} {\bibfnamefont {L.}~\bibnamefont
  {Hadji}},\ }\href@noop {} {\bibfield  {journal} {\bibinfo  {journal}
  {Physical Review E}\ }\textbf {\bibinfo {volume} {60}},\ \bibinfo {pages}
  {6180} (\bibinfo {year} {1999})}\BibitemShut {NoStop}%
\bibitem [{\citenamefont {Hadji}(2002)}]{Hadji2002}%
  \BibitemOpen
  \bibfield  {author} {\bibinfo {author} {\bibfnamefont {L.}~\bibnamefont
  {Hadji}},\ }\href {\doibase 10.1103/PhysRevE.66.041404} {\bibfield  {journal}
  {\bibinfo  {journal} {Physical Review E}\ }\textbf {\bibinfo {volume} {66}},\
  \bibinfo {pages} {1} (\bibinfo {year} {2002})}\BibitemShut {NoStop}%
\bibitem [{\citenamefont {Azouni}\ and\ \citenamefont
  {Casses}(1998)}]{Azouni1998}%
  \BibitemOpen
  \bibfield  {author} {\bibinfo {author} {\bibfnamefont {A.}~\bibnamefont
  {Azouni}}\ and\ \bibinfo {author} {\bibfnamefont {P.}~\bibnamefont
  {Casses}},\ }\href {\doibase 10.1016/S0001-8686(97)00002-X} {\bibfield
  {journal} {\bibinfo  {journal} {Advances in Colloid and Interface Science}\
  }\textbf {\bibinfo {volume} {75}},\ \bibinfo {pages} {83} (\bibinfo {year}
  {1998})}\BibitemShut {NoStop}%
\bibitem [{\citenamefont {Shangguan}\ \emph {et~al.}(1992)\citenamefont
  {Shangguan}, \citenamefont {Ahuja},\ and\ \citenamefont
  {Stefanescu}}]{Shangguan1992}%
  \BibitemOpen
  \bibfield  {author} {\bibinfo {author} {\bibfnamefont {D.}~\bibnamefont
  {Shangguan}}, \bibinfo {author} {\bibfnamefont {S.}~\bibnamefont {Ahuja}}, \
  and\ \bibinfo {author} {\bibfnamefont {D.~M.}\ \bibnamefont {Stefanescu}},\
  }\href {\doibase 10.1007/BF02801184} {\bibfield  {journal} {\bibinfo
  {journal} {Metallurgical Transactions A}\ }\textbf {\bibinfo {volume} {23}},\
  \bibinfo {pages} {669} (\bibinfo {year} {1992})}\BibitemShut {NoStop}%
\bibitem [{\citenamefont {Catalina}\ \emph {et~al.}(2000)\citenamefont
  {Catalina}, \citenamefont {Mukherjee},\ and\ \citenamefont
  {Stefanescu}}]{Catalina2000}%
  \BibitemOpen
  \bibfield  {author} {\bibinfo {author} {\bibfnamefont {A.}~\bibnamefont
  {Catalina}}, \bibinfo {author} {\bibfnamefont {S.}~\bibnamefont {Mukherjee}},
  \ and\ \bibinfo {author} {\bibfnamefont {D.}~\bibnamefont {Stefanescu}},\
  }\href {http://www.springerlink.com/index/6P8247636K2418Q8.pdf} {\bibfield
  {journal} {\bibinfo  {journal} {Metallurgical and Materials Transactions A}\
  }\textbf {\bibinfo {volume} {31}},\ \bibinfo {pages} {2559} (\bibinfo {year}
  {2000})}\BibitemShut {NoStop}%
\bibitem [{\citenamefont {Rempel}\ and\ \citenamefont
  {Worster}(1999)}]{Rempel1999}%
  \BibitemOpen
  \bibfield  {author} {\bibinfo {author} {\bibfnamefont {A.~W.}\ \bibnamefont
  {Rempel}}\ and\ \bibinfo {author} {\bibfnamefont {M.~G.}\ \bibnamefont
  {Worster}},\ }\href@noop {} {\bibfield  {journal} {\bibinfo  {journal}
  {Journal of Crystal Growth}\ }\textbf {\bibinfo {volume} {205}},\ \bibinfo
  {pages} {427} (\bibinfo {year} {1999})}\BibitemShut {NoStop}%
\bibitem [{\citenamefont {Rempel}\ and\ \citenamefont
  {Worster}(2001)}]{Rempel2001}%
  \BibitemOpen
  \bibfield  {author} {\bibinfo {author} {\bibfnamefont {A.~W.}\ \bibnamefont
  {Rempel}}\ and\ \bibinfo {author} {\bibfnamefont {M.~G.}\ \bibnamefont
  {Worster}},\ }\href {\doibase 10.1016/S0022-0248(01)00595-4} {\bibfield
  {journal} {\bibinfo  {journal} {Journal of Crystal Growth}\ }\textbf
  {\bibinfo {volume} {223}},\ \bibinfo {pages} {420} (\bibinfo {year}
  {2001})}\BibitemShut {NoStop}%
\bibitem [{\citenamefont {Garvin}(2003)}]{Garvin2003}%
  \BibitemOpen
  \bibfield  {author} {\bibinfo {author} {\bibfnamefont {J.}~\bibnamefont
  {Garvin}},\ }\href {\doibase 10.1016/S0022-0248(03)00943-6} {\bibfield
  {journal} {\bibinfo  {journal} {Journal of Crystal Growth}\ }\textbf
  {\bibinfo {volume} {252}},\ \bibinfo {pages} {467} (\bibinfo {year}
  {2003})}\BibitemShut {NoStop}%
\bibitem [{\citenamefont {Park}\ \emph {et~al.}(2006)\citenamefont {Park},
  \citenamefont {Golovin},\ and\ \citenamefont {Davis}}]{Park2006}%
  \BibitemOpen
  \bibfield  {author} {\bibinfo {author} {\bibfnamefont {M.~S.}\ \bibnamefont
  {Park}}, \bibinfo {author} {\bibfnamefont {A.~A.}\ \bibnamefont {Golovin}}, \
  and\ \bibinfo {author} {\bibfnamefont {S.~H.}\ \bibnamefont {Davis}},\ }\href
  {\doibase 10.1017/S0022112006000796} {\bibfield  {journal} {\bibinfo
  {journal} {Journal of Fluid Mechanics}\ }\textbf {\bibinfo {volume} {560}},\
  \bibinfo {pages} {415} (\bibinfo {year} {2006})}\BibitemShut {NoStop}%
\bibitem [{\citenamefont {Georgelin}\ and\ \citenamefont
  {Pocheau}(1998)}]{Georgelin1998}%
  \BibitemOpen
  \bibfield  {author} {\bibinfo {author} {\bibfnamefont {M.}~\bibnamefont
  {Georgelin}}\ and\ \bibinfo {author} {\bibfnamefont {A.}~\bibnamefont
  {Pocheau}},\ }\href@noop {} {\bibfield  {journal} {\bibinfo  {journal} {Phys.
  Rev. E}\ }\textbf {\bibinfo {volume} {57}},\ \bibinfo {pages} {3189}
  (\bibinfo {year} {1998})}\BibitemShut {NoStop}%
\bibitem [{\citenamefont {Pocheau}\ and\ \citenamefont
  {Georgelin}(1999)}]{Pocheau1999}%
  \BibitemOpen
  \bibfield  {author} {\bibinfo {author} {\bibfnamefont {A.}~\bibnamefont
  {Pocheau}}\ and\ \bibinfo {author} {\bibfnamefont {M.}~\bibnamefont
  {Georgelin}},\ }\href@noop {} {\bibfield  {journal} {\bibinfo  {journal} {J.
  Cryst. Growth}\ }\textbf {\bibinfo {volume} {206}},\ \bibinfo {pages} {215}
  (\bibinfo {year} {1999})}\BibitemShut {NoStop}%
\bibitem [{\citenamefont {Pocheau}\ \emph {et~al.}(2009)\citenamefont
  {Pocheau}, \citenamefont {Bodea},\ and\ \citenamefont
  {Georgelin}}]{Pocheau2009}%
  \BibitemOpen
  \bibfield  {author} {\bibinfo {author} {\bibfnamefont {A.}~\bibnamefont
  {Pocheau}}, \bibinfo {author} {\bibfnamefont {S.}~\bibnamefont {Bodea}}, \
  and\ \bibinfo {author} {\bibfnamefont {M.}~\bibnamefont {Georgelin}},\ }\href
  {\doibase 10.1103/PhysRevE.80.031601} {\bibfield  {journal} {\bibinfo
  {journal} {Phys. Rev. E}\ }\textbf {\bibinfo {volume} {80}},\ \bibinfo
  {pages} {031601} (\bibinfo {year} {2009})}\BibitemShut {NoStop}%
\bibitem [{\citenamefont {Deschamps}\ \emph {et~al.}(2008)\citenamefont
  {Deschamps}, \citenamefont {Georgelin},\ and\ \citenamefont
  {Pocheau}}]{Deschamps2008}%
  \BibitemOpen
  \bibfield  {author} {\bibinfo {author} {\bibfnamefont {J.}~\bibnamefont
  {Deschamps}}, \bibinfo {author} {\bibfnamefont {M.}~\bibnamefont
  {Georgelin}}, \ and\ \bibinfo {author} {\bibfnamefont {A.}~\bibnamefont
  {Pocheau}},\ }\href {\doibase 10.1103/PhysRevE.78.011605} {\bibfield
  {journal} {\bibinfo  {journal} {Physical Review E}\ }\textbf {\bibinfo
  {volume} {78}},\ \bibinfo {pages} {011605} (\bibinfo {year}
  {2008})}\BibitemShut {NoStop}%
\bibitem [{\citenamefont {Gurevich}\ \emph {et~al.}(2010)\citenamefont
  {Gurevich}, \citenamefont {Karma}, \citenamefont {Plapp},\ and\ \citenamefont
  {Trivedi}}]{Gurevich2010}%
  \BibitemOpen
  \bibfield  {author} {\bibinfo {author} {\bibfnamefont {S.}~\bibnamefont
  {Gurevich}}, \bibinfo {author} {\bibfnamefont {A.}~\bibnamefont {Karma}},
  \bibinfo {author} {\bibfnamefont {M.}~\bibnamefont {Plapp}}, \ and\ \bibinfo
  {author} {\bibfnamefont {R.}~\bibnamefont {Trivedi}},\ }\href {\doibase DOI:
  10.1103/PhysRevE.81.011603} {\bibfield  {journal} {\bibinfo  {journal} {Phys.
  Rev. E}\ }\textbf {\bibinfo {volume} {81}},\ \bibinfo {pages} {011603}
  (\bibinfo {year} {2010})}\BibitemShut {NoStop}%
\bibitem [{\citenamefont {Ghmadh}\ \emph {et~al.}(2014)\citenamefont {Ghmadh},
  \citenamefont {Debierre}, \citenamefont {Deschamps}, \citenamefont
  {Georgelin}, \citenamefont {Guerin},\ and\ \citenamefont
  {Pocheau}}]{Ghmadh2014}%
  \BibitemOpen
  \bibfield  {author} {\bibinfo {author} {\bibfnamefont {J.}~\bibnamefont
  {Ghmadh}}, \bibinfo {author} {\bibfnamefont {J.-M.}\ \bibnamefont
  {Debierre}}, \bibinfo {author} {\bibfnamefont {J.}~\bibnamefont {Deschamps}},
  \bibinfo {author} {\bibfnamefont {M.}~\bibnamefont {Georgelin}}, \bibinfo
  {author} {\bibfnamefont {R.}~\bibnamefont {Guerin}}, \ and\ \bibinfo {author}
  {\bibfnamefont {A.}~\bibnamefont {Pocheau}},\ }\href {\doibase
  10.1016/j.actamat.2014.04.023} {\bibfield  {journal} {\bibinfo  {journal}
  {Acta Materiala}\ }\textbf {\bibinfo {volume} {74}},\ \bibinfo {pages} {255}
  (\bibinfo {year} {2014})}\BibitemShut {NoStop}%
\bibitem [{\citenamefont {Anderson}\ and\ \citenamefont
  {Worster}(2012)}]{Anderson2012}%
  \BibitemOpen
  \bibfield  {author} {\bibinfo {author} {\bibfnamefont {A.~M.}\ \bibnamefont
  {Anderson}}\ and\ \bibinfo {author} {\bibfnamefont {M.~G.}\ \bibnamefont
  {Worster}},\ }\href {\doibase 10.1021/la303458m} {\bibfield  {journal}
  {\bibinfo  {journal} {Langmuir}\ }\textbf {\bibinfo {volume} {28}},\ \bibinfo
  {pages} {16512} (\bibinfo {year} {2012})}\BibitemShut {NoStop}%
\bibitem [{\citenamefont {Pieranski}(1983)}]{Pieranski1983}%
  \BibitemOpen
  \bibfield  {author} {\bibinfo {author} {\bibfnamefont {P.}~\bibnamefont
  {Pieranski}},\ }\href {\doibase 10.1080/00107518308227471} {\bibfield
  {journal} {\bibinfo  {journal} {Contemporary Physics}\ }\textbf {\bibinfo
  {volume} {24}},\ \bibinfo {pages} {25} (\bibinfo {year} {1983})},\ \Eprint
  {http://arxiv.org/abs/http://dx.doi.org/10.1080/00107518308227471}
  {http://dx.doi.org/10.1080/00107518308227471} \BibitemShut {NoStop}%
\bibitem [{\citenamefont {Peppin}\ \emph {et~al.}(2006)\citenamefont {Peppin},
  \citenamefont {Elliott},\ and\ \citenamefont {Worster}}]{Peppin2006}%
  \BibitemOpen
  \bibfield  {author} {\bibinfo {author} {\bibfnamefont {S.~S.}\ \bibnamefont
  {Peppin}}, \bibinfo {author} {\bibfnamefont {J.~A.~W.}\ \bibnamefont
  {Elliott}}, \ and\ \bibinfo {author} {\bibfnamefont {M.~G.}\ \bibnamefont
  {Worster}},\ }\href {\doibase 10.1017/S0022112006009268} {\bibfield
  {journal} {\bibinfo  {journal} {Journal of Fluid Mechanics}\ }\textbf
  {\bibinfo {volume} {554}},\ \bibinfo {pages} {147} (\bibinfo {year}
  {2006})}\BibitemShut {NoStop}%
\bibitem [{\citenamefont {Maijer}\ \emph {et~al.}(1978)\citenamefont {Maijer},
  \citenamefont {van Megen~W.J.},\ and\ \citenamefont {J.}}]{Meijer1978}%
  \BibitemOpen
  \bibfield  {author} {\bibinfo {author} {\bibfnamefont {A.}~\bibnamefont
  {Maijer}}, \bibinfo {author} {\bibnamefont {van Megen~W.J.}}, \ and\ \bibinfo
  {author} {\bibfnamefont {L.}~\bibnamefont {J.}},\ }\href@noop {} {\bibfield
  {journal} {\bibinfo  {journal} {Journal of Colloid Interface Sci.}\ }\textbf
  {\bibinfo {volume} {66}},\ \bibinfo {pages} {99} (\bibinfo {year}
  {1978})}\BibitemShut {NoStop}%
\bibitem [{\citenamefont {L.}\ \emph {et~al.}(2010)\citenamefont {L.},
  \citenamefont {W.J.},\ and\ \citenamefont {A.F.}}]{Goehring2010}%
  \BibitemOpen
  \bibfield  {author} {\bibinfo {author} {\bibfnamefont {G.}~\bibnamefont
  {L.}}, \bibinfo {author} {\bibfnamefont {C.}~\bibnamefont {W.J.}}, \ and\
  \bibinfo {author} {\bibfnamefont {R.}~\bibnamefont {A.F.}},\ }\href@noop {}
  {\bibfield  {journal} {\bibinfo  {journal} {Langmuir}\ }\textbf {\bibinfo
  {volume} {26}},\ \bibinfo {pages} {9269} (\bibinfo {year}
  {2010})}\BibitemShut {NoStop}%
\bibitem [{Note1()}]{Note1}%
  \BibitemOpen
  \bibinfo {note} {$\protect \mathaccentV {bar}016{P_{\protect \rm T}} \propto
  (\delta _0^{-2}+\delta _0^{-6})$ ; $f \propto (\delta _0^{-1}+ 4/5 \delta
  _0^{-5}+ 1/3 \delta _0^{-9})$, where $\delta _0$ stands for the
  non-dimensional film thickness $d_0/l$ with $l^4=\protect \mathcal {A}/6\pi
  \protect \tmspace +\thickmuskip {.2777em} T_M/L_v$.}\BibitemShut {Stop}%
\end{thebibliography}%

\end{document}